
%
%
\raggedbottom
\parskip=0.3cm
%

\magnification=\magstephalf
\vsize=9.0truein \voffset=-0.0truein
\hsize=6.5truein \hoffset=-0.0truein
\raggedbottom
\interlinepenalty=50
\baselineskip=14.0truept
\parskip=0.0truecm


%
%

\def\parano    {\par \noindent}

\def\noparskip {\parskip=0pt}
\def\lappeq{\lower 0.5ex\hbox{$\; \buildrel < \over \sim \;$}
               \allowbreak}
\def\gappeq{\lower 0.5ex\hbox{$\; \buildrel > \over \sim \;$}
               \allowbreak}
\def\etal {et al.\ }

\def\Tc   {$T_{c}$\ }
\def\Tco  {$T_{c}$}

\def\ct      {$^{12}$C\ }

\def\crat    {$^{13}$C/$^{12}$C\ }

\def\os      {$^{16}$O\ }

\def\orat    {$^{18}$O/$^{16}$O\ }
\def\net     {$^{20}$Ne\ }
\def\nett    {$^{22}$Ne\ }
\def\nerat   {$^{22}$Ne/$^{20}$Ne\ }
\def\heeeo   {$^{3}$He}

\def\cto     {$^{12}$C}

\def\oso     {$^{16}$O}

\def\neto    {$^{20}$Ne}
\def\netto   {$^{22}$Ne}
\def\nerato  {$^{22}$Ne/$^{20}$Ne}
\def\sol     {$_{\odot}$\ }
\def\solo    {$_{\odot}$}
%
%
%
\def\lesssim{\mathrel{\hbox{\rlap
  {\hbox{\lower4pt\hbox{$\sim$}}}\hbox{$<$}}}}
\def\gtrsim{\mathrel{\hbox{\rlap
  {\hbox{\lower4pt\hbox{$\sim$}}}\hbox{$>$}}}}
\def\eqb{\begin{equation}}
\def\eqe{\end{equation}}

\def\x#1{\times 10^{#1}}

\def\Heone{He$^{+1}$ }

\def\Hetwoo{He$^{+2}$}

\def\iec{i.e., }

\def\egc{e.g., }
\def\etal{et al.\ }

\newcount\listno
\listno= 0
\def\list{\global\advance \listno by 1 {(\the\listno) }}

\def\reff{\noindent \hangafter= 1 \hangindent= 0.5truecm}
%
%
%
%
\newcount\eqnno
\eqnno= 0
\newcount\eqnnoA
\eqnnoA= 0
\def\eqnum{\global\advance \eqnno by 1 \eqno{(\the\eqnno)}}
\def\eqnumA{\global\advance \eqnnoA by 1 \eqno{(A\the\eqnnoA)}}
\def\eqnumal{\global\advance \eqnno by 1   (\the\eqnno)}
\def\label#1{\global\advance \eqnno by 1\xdef#1{\the\eqnno }
    \global\advance \eqnno by -1}
\def\labelA#1{\global\advance \eqnnoA by 1\xdef#1{\the\eqnnoA }
    \global\advance \eqnnoA by -1}
%
%
%
%
\newcount\figno
\figno= 0
\def\fignum{\the\figno}
\def\labelfig#1{\global\advance \figno by 1 \xdef#1{\the\figno}}
%
%
%


%
%
%
\catcode`@=11
\newskip\ttglue
\font\eightrm=cmr9
\font\eighti=cmmi9
\font\eightsy=cmsy9
\font\eightbf=cmbx9
\font\eighttt=cmtt9
\font\eightit=cmti9
\font\eightsl=cmsl9
\font\sixrm=cmr7
\font\sixi=cmmi7
\font\sixsy=cmsy7
\font\sixbf=cmbx7
\font\cmmm=cmmi9
\textfont1=\cmmm
\def\eightpoint{\def\rm{\fam0\eightrm}
   \textfont0=\eightrm \scriptfont0=\sixrm \scriptscriptfont0=\fiverm
   \textfont1=\eighti  \scriptfont1=\sixi  \scriptscriptfont1=\fivei
   \textfont2=\eightsy \scriptfont2=\sixsy \scriptscriptfont2 \fivesy
   \textfont3=\tenex   \scriptfont3=\tenex \scriptscriptfont3=\tenex
   \textfont\itfam=\eightit  \def\it{\fam\itfam\eightit}%
   \textfont\slfam=\eightsl  \def\sl{\fam\slfam\eightsl}%
   \textfont\ttfam=\eighttt  \def\tt{\fam\ttfam\eighttt}%
   \textfont\bffam=\eightbf  \scriptfont\bffam=\sixbf
    \scriptscriptfont\bffam=\fivebf  \def\bf{\fam\bffam\eightbf}%
   \tt \ttglue=.5em plus.25em minus.15em
   \normalbaselineskip=9pt
   \setbox\strutbox=\hbox{\vrule height7pt depth2pt width0pt}%
   \let\sc=\sixrm  \let\big=\eightbig  \normalbaselines\rm}
\def\footnote#1{\edef\@sf{\spacefactor\the\spacefactor}#1\@sf
      \insert\footins\bgroup\eightpoint
      \interlinepenalty100 \let\par=\endgraf
        \leftskip=0pt \rightskip=0pt
        \splittopskip=10pt plus 1pt minus 1pt \floatingpenalty=20000
        \smallskip\item{#1}\bgroup\strut\aftergroup\@foot\let\next}
\skip\footins=12pt plus 2pt minus 4pt
\dimen\footins=30pc
%
%
%
\newcount\Inum
\newcount\IInum
\newcount\IIInum
\newcount\IVnum
\Inum=0
\def\I{\global\multiply\IInum by 0 \global\multiply\IIInum by 0
            \global\multiply\IVnum by 0 \global\advance \Inum by 1
            {\the\Inum. }}
\IInum=0
\def\II{\global\multiply\IIInum by 0\global\multiply\IVnum by 0
       \global\advance \IInum by 1 {\the\Inum.\the\IInum. }}
\IIInum=0
\def\III{\global\multiply\IVnum by 0\global\advance \IIInum by 1
            {\the\Inum.\the\IInum.\the\IIInum.}}
\IVnum=0
\def\IV{\global\advance \IVnum by 1
            {\the\IVnum.}}
\parindent=18truept

\def\cl#1{\centerline{#1}}
%
%
%
\count1=\number\time
\divide \count1 by 60
\def\hours{\number\count1}
\count2=\count1
\multiply\count2 by -60
\count3=\number\time
\advance \count3 by \count2
\def\minutes{\number\count3}
\def\timeexplicit{\hours:\minutes}
\def\today{\rm{\space\space\space\space\space\number\day\space
           \ifcase\month\or January\or February\or March\or April\or May\or
June
           \or July\or August\or September\or October\or November\or
December\fi
           \space\number\year\space\space\number\timeexplicit}}
%
%
%
%
%
%


\cl{\bf GALACTIC COSMIC RAYS FROM SUPERNOVA REMNANTS:}
\vskip3pt
\cl{\bf I \ A COSMIC RAY COMPOSITION CONTROLLED BY}
\vskip3pt
\cl{\bf VOLATILITY AND MASS-TO-CHARGE RATIO}

\vskip12pt
\cl{Jean-Paul Meyer}
\vskip4pt
\cl{Service d'Astrophysique, CEA/DSM/DAPNIA,
Centre d'Etudes de Saclay}
\cl{91191 Gif-sur-Yvette, FRANCE}
\cl{meyer@sapvxb.saclay.cea.fr}

\vskip12pt
\cl{Luke O'C. Drury}
\vskip4pt
\cl{Dublin Institute for Advanced Studies}
\cl{School of Cosmic Physics, 5 Merrion Square}
\cl{Dublin 2, IRELAND}
\cl{ld@cp.dias.ie}

\vskip12pt
\cl{Donald C. Ellison}
\vskip4pt
\cl{Department of Physics, North Carolina State University}
\cl{Box 8202, Raleigh NC 27695, U.S.A.}
\cl{don\_ellison@ncsu.edu}

\vskip18pt
\cl{Submitted to the {\it Astrophysical Journal,} February 10,  1997 }
\vskip12pt
\centerline{Accepted, April 5, 1997}

\vskip24pt

\cl{ABSTRACT}
\vskip8pt

We show that the Galactic Cosmic Ray source (GCRS) composition is best
described in terms of
(i) a general enhancement of the refractory elements relative to the
volatile ones, and
(ii) among the volatile elements, an enhancement of the heavier
elements relative to the lighter ones;
this mass dependence most likely reflects a mass-to-charge ($A/Q$)
dependence of the acceleration efficiency;
among the refractory elements, there is {\it no\/} such enhancement
of heavier species, or only a much weaker one.
We regard as coincidental the similarity between the GCRS composition
and that of the solar corona, which is biased according to first
ionization potential.
In a companion paper, this GCRS composition is interpreted in terms of
an acceleration by supernova shock waves of interstellar and/or
circumstellar (\egc \netto-rich Wolf-Rayet wind) gas-phase and
especially dust material.


\vskip12pt
\cl {1. INTRODUCTION}
\vskip8pt

The composition of Galactic Cosmic Rays (GCR) contains some of
the principal
clues regarding their origin.
In earlier times, it was believed that GCR's originate in newly
processed supernova (SN) ejecta, the SN explosion being held
responsible for both their general heavy element enhancement
relative to H and He, and their acceleration.
In the 70's, however, it became clear that the {\it detailed\/} GCR
source (GCRS) composition anomalies ({\S}~2.1) did not seem to be
controlled by nucleosynthetic processes ({\S}~2.2).  By contrast, they
seemed controlled by {\it atomic\/}, rather than nuclear, parameters,
such as the first ionization potential (FIP), the lower-FIP elements
being systematically in excess relative to the higher-FIP ones
({\S}~2.3).
This fact,
together with the finding of an {\it extremely similar\/} FIP-bias in
the composition of the solar corona, solar wind, and
solar energetic particles (SEP's),
and with the {\it lack\/} of a depletion in GCR's of the refractory
elements locked in grains in most of the interstellar medium (ISM),
suggested a cosmic ray  origin in the coronal material of
later-type stars possessing a cool, neutral H chromosphere in which an
ion-neutral separation could possibly take place.
Along this line, it has been conjectured that GCR's consisted of
stellar energetic particles with frozen-in coronal composition
(similar to SEP's), first injected at MeV energies
by stellar activity, and then later reaccelerated to GeV and TeV
energies by passing SNR shock waves ({\S}~2.3).
This view required an awkward two-stage acceleration process, in two
separate sites.
In addition, the presence of a \nett excess in GCR's further
required the presence of a totally unrelated second GCR component,
presumably originating in Wolf-Rayet wind material ({\S}~2.3).
This scenario also had difficulty accounting for the fairly large
spread of the enhancements among the high-FIP elements, and in
particular for the low abundances of H, He, and N.

It was, however, realized long ago that the FIP of the various
chemical elements and the volatility of the chemical compounds they
form are correlated:
typically, low-FIP elements (metals) form refractory compounds, while
high-FIP elements (hydrogen, non-metals, noble gases) form
volatile compounds or do not condense at all.  Therefore, the apparent
ordering of the GCRS composition in terms of FIP could as well reflect
an actual ordering in terms of volatility ({\S}~3)!
Such an ordering would imply an enhanced acceleration of those
elements locked in dust grains in the ISM, as compared to those in the
gas-phase.
Models for a preferential acceleration of ISM grain destruction
products by SNR shock waves were  actually explored in the early
80's ({\S}~3).

To remove the ambiguity and choose between FIP and volatility as the
relevant parameter, the behavior of those elements which are
exceptions to the above general correlation must be considered:
low-FIP volatile elements, and high-FIP refractories.  These elements
are not the easiest to observe in GCR's!  In {\S}~4, we find nine
such appropriate clue elements, whose GCRS abundance is reasonably
well determined.
Out of these, four are found clearly discrepant with FIP and
suggesting volatility as the relevant parameter.
The five other ones are consistent with both  FIP and volatility.

A tentative analysis of the data in terms of volatility, performed in
{\S}~5, shows that the GCRS abundances of {\it all\/} elements are
remarkably well organized in terms
%
of the combined effects of
(i) volatility, the more refractory elements being in excess relative
to the more volatile ones,
and (ii) mass, or more probably mass number-to-charge number ratio,
$A/Q$, in specific
ionization conditions, the more massive volatiles being in excess
relative to the less massive ones;
this description, in particular, accounts for the low GCRS abundances
of H, He and N, although the H/He ratio may be somewhat larger than
expected; this mass effect is much weaker or absent among the
refractory elements.
A similar combination of an ordering in terms of FIP and of mass
would not account for the data as satisfactorily.
Our conclusions are summarized in {\S}~6.

This behavior will be interpreted in a companion paper by Ellison,
Drury, \& Meyer (1997, Paper II), in terms of an $A/Q$ dependent
acceleration of interstellar and/or circumstellar volatile gas-phase
elements by smoothed SNR shock waves, and of a preferential
acceleration of entire dust grains followed by their sputtering,
accounting for the roughly mass-independent excess of refractory
elements.
In this scenario, the acceleration takes place in a single step and at
a single site; the \nett excess (Appendix) is also naturally accounted
for, since higher mass stars produce SNR shocks which accelerate their
own \netto-rich pre-SN Wolf-Rayet wind material.


\vskip12pt
\vbox {
\cl {2. GCR COMPOSITION: THE CURRENT GENERAL PICTURE}

\vskip12pt
\cl {2.1 {\it GCR Source Composition Determinations}}
\vskip8pt

The GCR source composition is derived from cosmic rays observed near
Earth by correcting for the effects of solar modulation, and of
spallation reactions and energy loss during interstellar propagation.
The deviations of the GCRS elemental composition relative to solar
abundances, assumed typical of our local galactic environment, have
been plotted versus element mass in Fig.~6 (and versus other
parameters in Figs.~1 and 5).
The GCRS overabundances relative to solar  are shown normalized
to H, for the best known energy range between $\sim$~1 to 30 GeV/n,
But the composition does not seem to change significantly at least up
to $\sim$~1000 GeV/n (except possibly for H, see below; \egc Shibata
1995).
The GCRS abundances for elements up to Ni are mainly based on the
HEAO-C2 abundances of Engelmann et al.\ (1990), the review by Ferrando
(1993), and the recent Ulysses data of Duvernois \& Thayer (1996);
those for the ``Ultra-Heavy" (UH) elements with $Z>$~30 on Binns
et al.\ (1989) and Binns (1995a);
more specific references for particular elements are given in {\S}~4.
The reference solar elemental abundances are taken from Grevesse,
Noels, \& Sauval (1996) (meteoritic determination adopted
preferentially), and the isotopic ratios from Anders \&
Grevesse (1989).
}  

Two remarks apply for UH elements.
First, the observations for $Z > 60$ have a limited charge resolution,
forcing us to deal only with {\it groups\/} of elements: especially
the ``Pt-group"  elements with $Z = 74$ -- 80 (hereafter ``Pt") and
the ``Pb-group" elements with $Z = 81$ -- 83 (hereafter ``Pb").
Second, the current data suggest GCRS excesses by factors of $\sim$~2
for many elements with $Z \gappeq$ 40 relative to Fe.
This applies, in particular, for the comparatively abundant ``Pt", as
well as for the secondaries with $Z\sim 61$ -- 73 produced by its
spallation.  By contrast, the rarer ``Pb" does {\it not\/} seem
enhanced relative to Fe.
Actually, in view of the increase of the total nuclear {\it
destruction\/} cross-sections with mass, the derived source abundances
of UH elements {\it relative to Fe\/} are very sensitive to the
propagation conditions;
(an excess of very short pathlengths in the GCR pathlength
distribution relative to the {generally assumed exponential
distribution [\egc Ptuskin \& Soutoul 1990]}, could
yield the observed excesses of UH elements, without excess of these
elements at the sources; this becomes more and more true for heavier
and heavier elements).
In addition, for $Z>60$, where only {\it groups\/} of elements can be
differentiated, errors on incompletely known, energy dependent, {\it
partial\/} cross sections also interfere in the derivation of the
source abundances relative to Fe.
So, the source ``Pt"/Fe and ``Pb"/Fe ratios cannot be precisely
determined.
The low source ``Pb"/``Pt" ratio is much better established, although
its value also depends on somewhat uncertain cross sections and on the
pathlength distribution (Binns et al.\ 1989; Clinton \& Waddington
1993; Waddington 1996, 1997); it is further discussed in {\S}~4.
In Figs.~1, 5, and 6, the error bars for those UH elements whose source
abundance {\it relative to Fe\/} might be affected by such poorly
known systematic errors are shown dashed and with a ``?" sign.

The abundances of dominant H and He relative to the heavier elements
deserve a special treatment.
We consider them in the same energy range where the heavy element
composition is best known, \iec mainly below $\sim$~30~GeV/n.
The source abundance of He has been assessed, based on Webber \&
McDonald (1994)'s observed low energy He/O ratios and their renewed
derivation of the source ratio, and on a comparison of the higher
energy He fluxes obtained mainly by Webber, Golden, \& Stephens (1987)
and Seo \etal (1991, 1995; and ref.\ therein) with Engelmann et al.\
(1990)'s observed O fluxes, extrapolated back to the source based on
Engelmann et al.\ (1985).
Altogether, we estimate the source He/O to lie in the range 19
$\pm$~4, \iec 0.145 $\pm$~0.030 times solar.
As for H, it has been anchored to the other elements through the
observed H/He ratio.
We considered mainly the data by Webber, Golden, \& Stephens (1987),
Seo \etal (1991, 1995; and ref.\ therein), Papini \etal (1993), and
Swordy
\etal (1995); see also Swordy (1994), and Shibata (1995).
Altogether, these data suggest a local interstellar H/He ratio around
23 $\pm$~5 at a given energy/nucleon in the $\sim$~5 to 30 GeV/n range
which, with a rigidity-dependent escape length from the Galaxy
$\propto R^{-0.6}$ applying above $\sim$~4~GeV/n, corresponds to a
$2^{0.6}$ times smaller source H/He ratio of 15 $\pm$~4
\footnote{$^1$}{If reacceleration of GCR's while propagating in
the galaxy is important, the composition data are accounted for by a
weaker rigidity dependence of the escape length ($\propto R^{-0.3}$);
however, the correcting factor relating the observed H/He ratio to the
source one would not be very different from the above one (Seo \&
Ptuskin 1994; Seo 1997).  Note that, with reacceleration, the derived
source spectra are softer than without, requiring a larger
contribution of weak shocks in the primary acceleration (Paper II).}.
There may be indications of some energy-dependence of the observed
H/He ratio within this range (the ratio seems much more constant at a
fixed rigidity), and of lower ratios at {\it much\/} higher energies
($\sim~10^{2}$ to $10^{4}$~GeV/n).
All in all, with (H/He)\sol = 10, H seems slightly enhanced relative to
He at a fixed energy/nucleon in the range we consider, by a factor of
1.5 $\pm$~0.4.

Key determinations of GCRS {\it isotopic\/} ratios will be found, \egc
in Leske (1993), Leske \& Wiedenbeck (1993), Ferrando (1993, 1994),
Lukasiak et al.\ (1994, 1997), Shibata (1995), Duvernois et al.\
(1996), and Connell \& Simpson (1997).


\vskip12pt
\vbox {
\cl {2.2 {\it Difficulties With the Interpretation of the GCR
              Composition}}
\cl      {\it in Terms of SN Nucleosynthesis}
\vskip8pt

For a long time, it was generally accepted that GCR's originate in
newly processed SN ejecta.
This view was very tantalizing, indeed, since GCR's are {\it
globally\/} enriched in heavy elements, while supernovae synthesize
heavy elements, disperse them, have ample energy available for
acceleration, and are actually observed to accelerate electrons.
It has, however, become clear that the {\it detailed\/} GCRS
composition
is inconsistent with a predominant selection of the elements according
to specific nucleosynthesis processes, and more particularly with what
could be expected from SN nucleosynthesis (Arnould 1984; Meyer 1985b,
1988).
}  

Examine the GCRS elemental composition anomalies presented versus
mass in Fig.~6.
First, considering heavy elements up to the Fe peak, $^{20}$Ne is found
depleted by a factor of $\sim 8$ relative to Mg, Al, and $\sim $ Na,
while these elements are all largely produced by C-burning.
Similarly, S and Ar are depleted by factors of $\sim 4$ relative to Si
and Ca, while these elements are all produced by O- and Si-burning
(\egc Meyer 1988, Fig.~2).  No such large anomalies are found in
existing SN nucleosynthesis calculations, especially between species
produced within the same burning cycle (Woosley \& Weaver 1995;
Timmes, Woosley \& Weaver 1995; Arnett 1995).
By contrast, Mg, Al (C-burning), Si and Ca (O- and Si-burning), and Fe
and Ni (e-process) are found to be very close to solar proportions in
GCRS's (within $\sim 20$\%), while the above nucleosynthesis
calculations for specific types of SNae commonly yield deviations of
these ratios by factors on the order of $\sim 2$.
%

Further, all GCRS isotope ratios are found consistent with solar, with
the important exception of the $^{22}$Ne/$^{20}$Ne and possibly the
$^{13}$C/$^{12}$C and $^{18}$O/$^{16}$O ratios, to be discussed later
(e.g.,\ Leske 1993; Leske \& Wiedenbeck 1993; Ferrando 1994; Lukasiak
et al.\ 1994, 1997; Duvernois et al.\ 1996; Connell \& Simpson
1997).
In particular, the isotopic $^{59}$Co/$^{57}$Co ratio indicates the
absence of freshly synthesized Fe peak nuclei (Leske 1993; Lukasiak et
al.\ 1997).

The composition of the GCR Ultra-Heavy (UH) elements with $Z>30$
largely confirms these views.
The observations of all elements through the first and the second $r$-
and $s$-process peaks (\iec up to $Z\sim 60$) show no trend for a
specific enhancement or deficiency of either $r$- or of $s$-nuclei.
The observations are roughly consistent with a solar source
composition affected by atomic selection effects discussed below
(possibly with the above discussed general excess of most elements
with $Z\gappeq40$ relative to Fe; Fig.~6)
(Binns et al.\ 1989; Clinton \& Waddington 1993; Binns 1995a,b;
Waddington 1996, 1997).
This observed {\it lack\/} of a $s$-element deficiency definitely
implies that GCR's do {\it not\/} predominantly originate in SN
processed material, since no type of SN synthesize $s$-nuclei
(Prantzos, Cass\'e, \& Vangioni-Flam 1993).
The apparent excess of the $r$-peak ``Pt", contrasting with the lack
of a $s$-peak ``Pb" excess, has been interpreted in terms of a
specific excess of the third $r$-process peak elements in the GCRS
material.  In view of the total absence of an excess of $r$-elements
in the first and second $r$-peaks, this hypothesis doesn't seem very
likely.  The low GCRS ``Pb"/``Pt" ratio probably requires another
explanation.

In brief, the current state of the art suggests that the GCRS
abundances of most elements is not controlled by specific
nucleosynthesis processes, and in particular not by SN
nucleosynthesis.
In fact, it seems that most of the GCR source material is more of the
``solar mix" type.
There is, however, an important exception to this statement: GCR
sources are enriched in the isotope \netto, suggesting the presence of
a He-burning material component most likely enriched in \ct and \os
as well as \netto, presumably originating in Wolf-Rayet star wind
material; this will be discussed in {\S}{\S}~ 2.3, 3, 5, 6, and in the
Appendix.


\vskip12pt
\vbox {
\cl {2.3 {\it The ``FIP" Plus He-Burning, Later-Type-Stars Plus
Wolf-Rayet Picture}}
\vskip8pt

Early in the `70s, it was noted that the {\it detailed\/} GCRS
heavy element composition, while not easily ordered in terms of nuclear
physics parameters, could be rather well organized in terms of atomic
physics parameters.
The FIP, or related parameters which control the tendency of an
element to be neutral or ionized in a gas at $\sim 10^4$~K (or
subjected to a radiation of comparable energy), seemed to roughly order
the deviations of the GCRS elemental composition relative to solar
abundances, at least for elements up to Ni
(Cass\'e \& Goret 1978 and ref.\ therein; Meyer 1985b; Arnaud \&
Cass\'e 1985; Silberberg \& Tsao 1990).
As shown in Fig.~1, heavy elements with FIP$\lappeq 8.5$~eV
(``low-FIP") are typically enhanced by a factor of $\sim 5$ relative
to those with FIP$\gappeq~11$~eV (``high-FIP").
}  

The low temperature required for the parent gas ($\sim 10^4$~K),
together with the lack of a depletion in GCR's of the refractory
elements locked in grains in virtually all but the hottest
interstellar medium (Cass\'e, Goret \& Cesarsky 1975; Cass\'e \&
Goret 1978; Dwek 1979), first suggested that the GCR nuclei did not
originate in the ISM, but in stellar surfaces (Meyer, Cass\'e, \&
Reeves 1979; Meyer 1985b).
This conclusion, of course, assumed implicitly that only gas-phase
atoms of the ISM could be accelerated.

A concomitant advance in a totally different context, our own Solar
environment, then strongly influenced our views on the GCR source
material.  Hints were first found in the observed gradual event SEP
compositions, suggesting FIP-related anomalies, much like those found
in GCR's.  The difficulty here, was that the composition of SEP's is
changing all the time.  Meyer (1985a) and Breneman \& Stone (1985)
managed to separate out the permanent FIP-bias imprint on the data,
clearly related to the composition of the heliospheric source
material, from the rigidity-dependent variations of the composition
resulting from variable conditions of particle acceleration.  The
Solar Wind and the spectroscopic studies of the Solar coronal gas
(EUV, X-rays, nuclear $\gamma$-rays) have progressively confirmed the
presence of FIP-biased gas in the entire corona and outer heliosphere
outside coronal holes (\egc Meyer 1985b, 1993, 1996; Feldman 1992;
Geiss, Gloeckler, \& von Steiger 1995; Garrard \& Stone 1993; Reames
1995).
%
The similarity between the
GCRS and the Solar coronal composition, therefore, strongly supported the
earlier arguments suggesting that the GCR nuclei had been first extracted
from stellar atmospheres.

The parent gas of the GCR particles could now be specified more
precisely: probably the coronae of F to M later-type stars
possessing, like the Sun, a cool, predominantly neutral-H chromosphere
at around $\sim 7000$~K, in which, in some
yet debated way,
ionized heavies may be separated from neutral ones, and rise
preferentially into the corona (Meyer 1985b; review by H\'enoux
1995).  Efforts have been recently devoted to try to observe the
FIP-effect in the coronae of later type stars, thanks to the
instruments on board the EUVE and ASCA spacecraft, with variable
outcomes (Laming, Drake, \& Widing 1996; Drake, Laming, \& Widing
1997; Singh, White \& Drake 1996; and ref.\ therein).

Stellar flare activity is, however, certainly energetically unable
to accelerate the bulk of the (GeV) GCR's in the galaxy.
It may, however, accelerate some of the ``FIP-biased" coronal material
to low (MeV) energies, just as the Sun accelerates SEP's, thus
providing a suprathermal component with ``frozen" coronal composition.
These ``injected" MeV particles must then be later {\it
preferentially\/} accelerated by more powerful SNR shock waves, which
boost them to the GeV and TeV energies of GCR's (Meyer 1985b).  So,
this scenario requires two separate acceleration stages in two
separate sites, clearly an undesirable feature.

Further, later-type star coronal gas cannot be the source for the
\netto -rich material required to account for the observed GCRS \nett
excess.
This \nett excess, the {\it only\/} significant source isotopic
anomaly found besides possibly low \crat and \orat ratios, together
with the high GCRS elemental C/O ratio, suggests the presence of a
pure He-burning material component in GCR's, whose most likely origin
is Wolf-Rayet wind material (Appendix).
So, another weak point of the FIP/later-type-star scenario is that an
additional, entirely unrelated source, is required for the
\netto-C-O-rich components.

Fig.~1 shows an up-to-date version of the correlation of the GCRS
abundance enhancements with FIP.
Obviously, FIP does roughly order the data.  But there is a lot of
scatter around the correlation.
Among the high-FIP elements, H and He are deficient relative to all
heavies, as has been known early on.
Regarding N, accurate GCR isotopic observations and spallation
cross-sections have now well established its low source abundance.
We have plotted the points for C and O as upper limits, because we
expect specific \ct and \os contributions associated with the \netto
-rich component from Wolf-Rayet stars, discussed in the Appendix; for
\cto, we propose a tentative estimate of its {\it non\/}-Wolf-Rayet
source abundance (Appendix).
%
Other elements deviate from the correlation.  Among lower-Z elements,
the source abundance of Na seems low, and that of P high, at least
based on the currently best available spallation cross sections for
these largely secondary species
(see discussion in {\S}~4).
The UH element data by and large confirm the FIP correlation, but with
a larger scatter, and a general trend towards larger enhancements for
$Z\gappeq 40$.  This trend may be real, or due to an improper account
of the interstellar propagation ({\S}~2.1).  An important exception is
Ge, which is reliably determined to be low as compared to Fe, with
exactly the same FIP value ({\S}~4).
Finally, while the ``Pt" and ``Pb" abundances relative to Fe are poorly
determined ({\S}~2.1), the low ``Pb"/``Pt" ratio also conflicts with
a FIP ordering, as also illustrated in Fig.~1.

\vskip12pt
\vbox {
\cl {3.  THE FIP VERSUS VOLATILITY ISSUE, AND GRAIN ACCELERATION
MODELS}
\vskip8pt

There exists for most elements a general correlation between the FIP
of each element and the volatility of the chemical compounds it forms.
This is illustrated in Fig.~2, which shows a cross plot of the
element condensation temperature \Tc versus its FIP, for all elements
(updated from Meyer 1981b).
This temperature \Tc is the calculated 50\% condensation temperature
in a $10^{-4}$~atm gas with solar composition, taken from Wasson
(1985).  The lower \Tco, the higher the volatility of the element.
Fig.~2 shows that FIP and \Tc are, indeed, anti-correlated for the
majority of the elements.
}  

Therefore, the apparent correlation of the GCRS abundances
with FIP could just {\it mimic} an actual correlation with the element
volatility.
With this viewpoint, the refractory (low-FIP!)  elements, those
generally locked in grains in the ISM, would be overabundant in GCR's.
This would imply {\it a preferential acceleration of grain destruction
products\/}, presumably in SNR shock waves.
One nice point with such a scenario is that the same SNR shock
waves could destroy the grains and fully accelerate the particles to
their final GeV and TeV energies.  We are back to a one-step
acceleration process, in a single site.

Now, it is believed that SNR shocks accelerate mainly external
interstellar or circumstellar material, not the SN ejecta themselves.
As discussed in Paper II, the role of the outer, forward shock should
indeed be dominant, the reverse shock being less energetic and short
lived, so that any particles it may accelerate later suffer severe
adiabatic losses (Drury \& Keane 1995).
Further, while instabilities in the flow may allow some of the ejecta
material to speed ahead of the external shock, this effect is believed
to be comparatively minor (\egc Jun \& Norman 1996; Drury \& Keane
1995).
It is therefore no surprise that we find little trace of SN
nucleosynthesis in the GCRS composition ({\S}~2.2), in spite of the
key role played by SNR shocks in accelerating the particles!
This applies to all types of SN, Type~I as well as Type~II.

The next important question is, of course: what does this external
material consist of?
Around Type~I SNae and lower mass core collapse SNae, i.e.\ Type~II's
with comparatively weak winds prior to explosion, this material ought
to be ordinary interstellar material (ISM), with roughly solar
composition (gas + grain).
Its grains consist of old ISM grains.  As one proceeds to more and
more massive SN progenitors, earlier wind ejections may become more
and more important, so that the shocks may accelerate the progenitor's
own wind material.  Its grains will then consist of newly formed
grains, presumably with different properties.
This is almost certainly the case for the highest mass, Wolf Rayet
(WR) star progenitors, which have been stripped off by huge winds, to
the point where their He-burning layers have been tapped and their
winds are highly enriched in He-burning products (\egc Van de Hucht \&
Hidayat 1991; Van de Hucht \& Williams 1995).
So, another extremely nice feature of this type of scenario is
that it may {\it account in a natural, continuous way} for the observed
excess of He-burning material ($^{\rm 22\/}$Ne, $^{\rm 12\/}$C, possibly
$^{\rm 16\/}$O excesses) in GCR sources.

Note that, in this context, the logic of the ``grain constraint"
earlier put forward by Meyer, Cass\'e, \& Reeves (1979) to {\it
exclude\/} the ordinary ISM as a possible source of the GCR ions, is
completely reversed: the ISM was excluded as a possible source of the
GCR's, based on the implicit assumption that ions had to be
accelerated out of the ISM {\it gas-phase\/}, which is depleted of its
low-FIP, refractory elements locked in grains.  Here we turn the
argument around, considering the opposite possibility of a
preferential acceleration of this very material locked in grains!

Such a preferential acceleration of grain material in SNR shock waves
was considered in the early 1980's.  Epstein (1980) first introduced
the concept considered in the present work: an acceleration of the
entire grain, followed by grain sputtering and by a re-acceleration of
the suprathermal grain destruction products to GeV and TeV energies.
Cesarsky \& Bibring (1981) and Bibring \& Cesarsky (1981), on the
other hand, suggested that grains freely cross the shock, so that they
acquire a bulk speed relative to the ambient shocked gas equal to the
shock speed.
When these grains undergo strong sputtering in the downstream region,
the sputtered ions are thus injected with a speed equal to the shock
speed, relative to the ambient gas, and preferentially {\it
stochastically\/} accelerated to cosmic ray energies.  First-order
Fermi acceleration in the shock itself is, indeed, unlikely for these
particles in this scenario, because most ions are sputtered off too
far downstream of the shock to diffuse back to it.
Of these two ideas, we believe Epstein's is most likely because of the
well-known problems with stochastic acceleration of cosmic rays.
First, for stochastic acceleration to be efficient requires Alfv\'en
Mach numbers considerably lower than those expected for
most remnants (e.g., Reynolds and Ellison 1993).
Secondly, unlike first-order Fermi acceleration, different ion species
(as well as the same species in different environments) can acquire
quite different spectral shapes (e.g., Forman,
Ramaty, \& Zweibel 1986), leaving one of the basic observations, i.e.,
that all cosmic ray ions have similar power laws, unexplained at a
fundamental level.
%

Further, Meyer (1981b), Tarafdar \& Apparao (1981), Soutoul et
al.\ (1991), and Sakurai (1995 and ref.\ therein), have provided
analyses of the GCRS composition data in the light of the condensation
of elements into grains.

Before we proceed to analyze the GCR data, we review the significance
of the condensation temperature \Tco, and its limitations.
We examine to what  point the actual, observed
composition of two types of astrophysical condensed materials seems
organized in terms of \Tco.
We first consider carbonaceous chondrite (CC) meteorites in the Solar
System.
It is now well established, by comparison with spectroscopic solar
photospheric abundance determinations, that in Type~1 CC's (C1) all
elements except H, C, N, O, and the noble gases, however volatile, are
condensed in their original proportions in the protosolar material
(e.g., Wasson \& Kallemeyn 1988; Anders \& Grevesse 1989; Grevesse
et al.\ 1996).
This means that C1's have not gone through any significant heating
phase.
In Type~2 and 3 CC's, by contrast, larger and larger fractions of the
material did go through hot phases, so that they are more and more
depleted of their more volatile elements, relative to C1's.
Fig.~3 illustrates this depletion for C3's, by showing the C3/C1
abundance ratio versus \Tco, for all elements entirely condensed in
C1's.
It can be seen that \Tc is a relevant parameter in organizing the
C3/C1 abundance ratios.

On the basis of Fig.~3, we define four groups of
elements: (i)\ ``refractory" elements with \Tc $>$ 1250~K, including
the numerous very refractory metals and the Mg, Si, Fe group
condensing in silicates and metallic Fe, with C3/C1 ratios
$\gappeq$~0.6; (ii)\ ``semi-volatile" elements with 1250~K $>$ \Tc $>$
875~K, with quite \Tco-dependent C3/C1 ratios between $\sim$~0.5 and
$\sim$~0.3; (iii)\ ``volatile" elements with 875~K $>$ \Tc $>$ 400~K,
with C3/C1 ratios $\lappeq$~0.3; and (iv)\ ``highly-volatile" elements
with 400~K $>$ \Tco, which are not significantly condensed even in
C1's, and include H, C, N, O, and noble gases.

While ``highly volatile" H and noble gases can in no case be
significantly trapped in solid bodies, C, N, and O may, in some cases
be partly condensed in silicates and oxides, solid carbons, organic
grain mantles, and fragile ices (e.g., 7\% of C, 1\% of N, and 38\%
of O are condensed in C1's).
But it seems very unlikely that they are significantly condensed in a
medium with solar composition in which ``volatile" elements with somewhat
higher \Tc values, such as, e.g., S, are still in the gas phase.
However, a very significant fraction of C may be condensed in solid
carbons if condensed in a C-rich atmosphere (C/O~$>$~1), such as Red
Giants and WC Wolf-Rayet wind material.
Carbon, indeed, behaves as a highly volatile element when entirely
trapped in gaseous CO molecules at high temperature; this is the case
whenever C/O~$<$~1, and in particular for a solar composition (C/O =
0.48).  But it behaves as a highly refractory element when CO
formation is hindered for lack of a sufficient amount of O in the
medium (C/O~$>$~1).
This remark may be important when interpreting the C (and O) excess
relative to the GCRS/Solar vs.\ $A$ correlation (Fig.~6), which might
have to be interpreted in terms of, not only Wolf-Rayet star
nucleosynthesis, but also of some of the C (O) preferentially
accelerated from grain material ({\S}~5).

A second context in which we can test the relevance of \Tc to the
fraction of condensed material is the observed depletion of the more
refractory elements in the cold ISM gas phase (e.g., Cardelli 1994;
Savage \& Sembach 1996).
It is illustrated in Fig.~4, showing the elemental depletions
relative to solar abundances along the particularly well studied line
of sight of $\zeta$ Oph.
The general trend for an increasing condensation in grains for
increasing \Tc is clear, in spite of a very large spread of the
depletions for any given \Tco, and of possible problems with the solar
abundance normalization (e.g., Mathis 1996; Dwek 1997).
There are also large differences from one line of sight to another,
but with always the same general trend.
The large spread of the depletions for a given \Tc might result from a
slow chemical reprocessing governing grain destruction and growth in
the ISM, largely independent of \Tco,
subsequent to the primary grain condensation phase in cooling stellar
ejecta and winds, which could be more closely controlled by \Tco;
a major grain regrowth in the ISM seems, indeed, required, in view of
the short lifetime of each individual grain in the ISM
(e.g., Joseph 1988; Draine 1990; Cardelli 1994; Savage \& Sembach 1996
and ref.\ therein).
In that sense, newly formed grains in stellar wind envelopes might
have a composition different from old ISM grains (Fig.~4), and
more closely controlled by \Tco.
All this may, in particular, apply for the comparatively refractory
elements P and As ($T_{c}\sim 1150$~K), which are found to be only
slightly depleted in the ISM (Fig.~4); they might be more locked in
grains in circumstellar material than in the ISM.


\vskip12pt
\vbox {
\cl {4. CLUES SUGGESTING THE RELEVANCE OF VOLATILITY}
\cl    {TO THE GCRS COMPOSITION}

\vskip8pt
We now ask the question: are there observational clues, which will
allow us to choose between FIP and volatility as the key factor
governing the GCRS abundances?  This question was
addressed early on by  Meyer (1981b), and we now update that analysis.
}  

To investigate this point, let us look closer into the correlation
between FIP and volatility for the various elements, shown in
Fig.~2.  To distinguish between the two types of scenarios, in
terms of FIP or of volatility, we have to look at the GCRS abundances
of those few elements which do {\it not\/} fit into the general
correlation between FIP and volatility (Meyer 1981b):
(i)\  low-FIP volatile elements, which should have solar abundances
relative to the standard, refractory low-FIP elements if FIP is
relevant, and should be comparatively depleted if \Tc is relevant;
(ii)\ high-FIP refractory elements, which should be depleted relative
to these same low-FIP elements if FIP is relevant, and should have
solar abundances if \Tc is relevant.

In Fig.~2, we have singled out those elements for which we
currently have reasonably accurate GCRS abundances (large, solid
dots).  Among them, $_{11}$Na, $_{15}$P, $_{16}$S, $_{29}$Cu,
$_{30}$Zn, $_{31}$Ga, $_{32}$Ge, $_{34}$Se, $_{82}$Pb lie outside or
only marginally within the FIP -- \Tc correlation, and are therefore
elements of interest in this context (framed in Fig.~2).
Unfortunately, these elements are not those for which the GCRS
abundance is easiest to determine!  We now investigate each of them,
in turn:

\vskip 0.3truecm \noindent {\it Sodium (Z = 11):\/}
In spite if its very low FIP of 5.1~eV, Na is a rather volatile
element ($T_{c}=970$~K; Fig.~3).
Na seems deficient by a factor of 2.0 $\pm$~0.8 relative to Si
(Ferrando 1993; DuVernois \& Thayer 1996).
This points towards volatility as the relevant parameter, controlling
the GCRS composition.
The measured Na abundance in arriving GCR's is absolutely foolproof
(Engelmann et al.\ 1990; Duvernois, Simpson, \& Thayer 1996).
However, Na is a predominantly secondary element in these observed
GCR's, so that the determination of its source abundance is very
sensitive to spallation cross-section errors; while the large error
bar on its adopted source abundance is based on a conservative
estimate of these errors, it might still be not entirely definitive.

\vskip 0.3truecm \noindent {\it Sulphur (Z = 16), Zinc (Z = 30), and
Selenium (Z = 34):\/}
These three elements form a group (Fig.~2).  All three have
neighboring values of FIP (10.4, 9.4, and 9.8~eV) which place them in
the ``intermediate-FIP" region where the amplitude of the FIP-bias
seems to be a rapidly varying function of the FIP value.
All three also have very similar values of \Tc (650, 660, and 680~K),
which make them full-fledged volatile elements (Fig.~3).
The source abundances of those elements are reliably determined.
Sulphur is one of the best known elements in GCR's (\egc Engelmann
\etal 1990).
We have good data on Zn from a clean pre-HEAO-C3 balloon flight, and
from both the C2 and the C3 instruments on board the HEAO-3
spacecraft, which converge on a source Zn/Fe ratio of 0.43 $\pm$~0.08
times solar (Tueller et al.\ 1979; Lund 1984; Binns et al.\ 1981;
Binns 1995a; Israel 1996).
Finally, Se is well measured by both the Ariel and the HEAO-C3
instruments (Fowler \etal 1987; Binns \etal 1989; Binns 1995a).
The secondary fraction of all three elements is small, and the
destruction cross sections of Zn and Se do not differ much from that
of Fe, so that the source Zn/Fe and Se/Fe ratios are close to
the observed ones, with a minor error due to interstellar propagation.
The GCRS abundances of these three elements is well interpreted in
terms of FIP (Fig.~1).  In terms of volatility, the lower GCRS
abundance of S, as compared to Zn and Se, may, at first, seem
disturbing; it will later be interpreted as a mass effect.

\vskip 0.3truecm \noindent {\it Phosphorus (Z = 15):\/}
The FIP of P, another ``intermediate-FIP" element, is 10.5~eV, i.e.\
virtually the same as that of S (10.4~eV).
But, while S is a full-fledged volatile ($T_{c}=650$~K), P is a rather
refractory semi-volatile element ($T_{c}=1150$~K; Fig.~3).
While S is depleted by a factor of $ 3.4 \pm 0.5$ relative to Si, P
seems depleted by a factor of 1.5 $\pm$~0.7 only (consistent with being
un-depleted), so that the P/S ratio is enhanced relative to solar by a
factor of 3.0 $\pm$~1.6 (Ferrando 1993; DuVernois \& Thayer 1996; see
also Leske \& Wiedenbeck 1993, and Duvernois, Simpson, \& Thayer
1996).
Thus, the high P/S ratio represents another hint in favor of volatility
controlling the GCRS composition.
There are, however, two caveats.
First, like Na, P is a predominantly secondary element in
the observed GCR's, so that the determination of its source abundance
is very sensitive to spallation cross-section errors.
Second, while P is a clearly siderophile element and does
behave as a rather refractory element in CC's, where the fractionation
seems well controlled by \Tc (Fig.~3), it seems surprisingly little
depleted in the ISM gas-phase, much less than other elements with
comparable values of \Tco, and actually hardly more than S (Fig.~4).
As discussed in {\S}~3, this difference might be due to the slow
chemical reprocessing of the grains in the ISM;
if this were the case, P could be much more condensed in the grains
recently formed in pre-SN stellar winds than in the general ISM
depicted in Fig.~4.

\vskip 0.3truecm \noindent {\it Copper (Z = 29) and Gallium (Z = 31):\/}
Cu and Ga are low-FIP (7.7 and 6.0~eV), semi-volatile elements
(\Tc = 1040 and 920~K; Fig.~3).
We have, unfortunately, only one observation of these odd-Z elements
in GCR's, by the HEAO-C2 experiment, in which these elements seem well
resolved (Byrnack et al.\ 1983; Lund 1984).
It yields Cu/Fe = 1.14 $\pm$~0.25 and Ga/Fe = 1.51 $\pm$~0.59 times
solar.  These values are consistent with FIP as the relevant
parameter.

\vskip 0.3truecm \noindent {\it Germanium (Z = 32):\/}
Ge has virtually the same FIP as Fe (7.9 eV), but Ge is a volatile
element ($T_{c}=825$~K), while Fe is refractory ($T_{c}=1135$~K;
Fig.~3).
The C2 and the C3 instruments on board the HEAO-3 spacecraft have
yielded independent, consistent GCRS Ge/Fe ratios, both significantly
lower than solar.  All in all, they lead to a GCRS Ge/Fe = 0.57
$\pm$~0.10 times solar (Lund 1984; Binns et al.\ 1989; Garrard et al.\
1990; Binns 1995a; Israel 1996).
In both instruments, the charge resolution is appropriate to safely
observe Ge, and possible systematic errors are limited.  The errors are
predominantly statistical.
As for Zn and Se, the {\it source\/} Ge/Fe ratio is close to the
measured one (only some $\sim 13$\% below; Binns et al.\ 1989), and
the error due to interstellar propagation is insignificant.
Therefore,
these considerations of Ge strongly argue in favor of volatility
controlling the GCRS source composition.

\vskip 0.3truecm \noindent {\it Lead (Z = 82):\/}
As discussed in {\S}~2.1, we will be comparing the abundances of the
``Pb" elements with $Z = 81$ -- 83, essentially made of low-FIP,
volatile, $s$-elements, with those of the ``Pt" elements with $Z = 74$
-- 80, mostly made of intermediate-FIP, refractory, $r$-elements
(Figs.~2 and 3).
While the observations suggest an excess of most elements with $Z
\gappeq 40$ relative to Fe at the sources,
the derived source abundances {\it relative to Fe\/} are very
sensitive to the propagation conditions  ({\S}~2.1; Fig.~1).
Here we will therefore deal only with the ``Pb"/``Pt" ratio, which is
much less affected by interstellar propagation.
According to standard calculations, the source ``Pb"/``Pt" ratio is
estimated to be $\sim 1.65$ times lower than the observed one (Binns
et al.\ 1989).  Depending upon the propagation conditions
(distribution of short pathlengths, {\S}~2.1), the adopted partial
cross-sections, and the source abundances themselves, this factor
could actually lie anywhere between
$\sim 1.3$ and $\sim 2.6$
(e.g., Clinton \& Waddington 1993; Waddington 1996, 1997).
Our current knowledge of the ``Pb"/``Pt" ratio comes essentially from
the Ariel-6 and the HEAO-C3 spacecraft experiments (Fowler et al.\ 1987;
Binns et al.\ 1989; Binns 1995a).
Recent experiments on board the LDEF facility currently yield
only very preliminary results, which do not conflict with the earlier
ones (O'Sullivan et al.\ 1995; Tylka et al.\ 1995; Domingo et al.\
1995).
Both available sets of data yield low ``Pb"/``Pt" ratios, altogether
consistent with an observed ``Pb"/``Pt" ratio
$\sim$~3.9 $\pm$~1.1
times lower than solar, which results in a source ``Pb"/``Pt" ratio
$\sim$~2.4 $\pm$~1.3
times lower than solar.
``Pb" elements are all low-FIP elements ($\sim 7.4$~eV), while ``Pt"
ones are mostly intermediate-FIP elements ($\sim 9$~eV).  Based on a plain
FIP-biased solar source composition (Fig.~1), one would therefore
expect a source ``Pb"/``Pt" ratio slightly higher than solar, by a
factor of
$\sim$~1.6.
So, the actual source ``Pb"/``Pt" ratio is
$\sim$~3.9 $\pm$~2.0
times lower than would be expected, based on a FIP-biased solar source
composition
(see Fig.~1).
This low ratio has been interpreted in terms of an excess of the {\it
third r}-peak Pt-group elements in the GCR sources ({\S}~2.2).
But it seems very difficult to have an excess of the {\it third
r\/}-peak nuclei without any excess of the lighter {\it r\/}-nuclei
({\S}~2.2).
The other possible interpretation is that ``Pb" is depleted relative to
``Pt" because ``Pb" elements are very volatile (\Tc $\sim 500$ K) while
``Pt" elements are refractory (\Tc $\sim 1400$ to 1800 K; Fig.~3).

\vskip 0.3truecm

In brief, the following picture emerges from this analysis
(Fig.~1):
\vskip 0.3truecm
{\parskip = 0pt
\parano -- One very solid indicator, Ge, and three less foolproof, but
still significant, indicators, Na, P, and
Pb point to volatility, not FIP,
as the relevant parameter governing the GCRS composition.
\parano -- The other five indicators, S, Zn, Se, Cu, and
Ga are consistent
with either the FIP picture or volatility.
In terms of volatility, the low S/Zn,Se ratio
may seem problematic, and semi-volatile Cu and Ga seem rather high.
In {\S}~5, these apparent difficulties will be interpreted in terms of
a mass effect.
} %


\vskip12pt
\vbox {
\cl {5. GCRS COMPOSITION: AN INTERPRETATION }
\cl {IN TERMS OF VOLATILITY AND MASS-TO-CHARGE RATIOS}
\vskip8pt

The outcome of the above analysis of the clue elements is sufficiently
suggestive to warrant a plot of the same GCRS overabundances relative
to Solar as in Fig.~1, but this time versus \Tc (Fig.~5).
For the ``highly volatile" elements with $T_{c}<400$~K, \Tc has
no physical relevance, and we have just ordered these
elements by mass.  Two conclusions can be drawn from Fig.~5:
\vskip 0.3truecm
{\noparskip
\parano -- By and large, the enhancement of the refractory elements
relative to the volatile ones is obvious.  More specifically, the two
``semi-volatile" and ``volatile" intermediate classes do tend to
show intermediate overabundances, but with quite a large scatter.
\parano -- By and large, the overabundances of the ``highly volatile"
elements seem an increasing function of their mass.  One exception is
H, which is slightly enhanced relative to He at a given
ener\-gy/nu\-cle\-on.
Further, C and O do not follow the trend.  But these are
precisely the two elements for which we expect a specific contribution
from WR stars.
In the Appendix, this WR contribution has been roughly estimated for
C, based on the WR nucleosynthetic yields only, \iec assuming that all
the C lies in the gas-phase.
We have not considered a possible preferential acceleration of a
significant fraction of the C which is locked in the grains formed in
the C-rich WC wind material in which C/O~$>$~1~ (see {\S}~3).
So, our assessment of the WR C contribution may be an underestimate;
that of the non-WR C abundance, plotted in Fig.~5, may therefore be
an overestimate.
} 
}  
\vskip 0.3truecm

This leads us to suspect that the overabundances of the elements in
the other classes of volatility might also be correlated with mass.
We therefore plot in Fig.~6 the same overabundance versus element
mass, distinguishing the elements in the four classes of volatility.
Fig.~6 contains the essential conclusions of this paper:
\vskip 0.3truecm
{\noparskip
\parano -- The overabundances of most ``highly volatile" elements are a
strongly increasing function of their atomic mass number, roughly
going as $\propto A^{0.8\pm0.2}$.
\parano -- C and O, the two very elements for which we expect a
specific contribution from WR stars (Appendix), are totally out of the
correlation.  As discussed above, we give in Fig.~6 an assessment
of the non-WR C abundance, which may still be an overestimate.  It
agrees reasonably well with the trend given by the neighboring
elements.
\parano -- H is also less depleted than expected based on the pattern for
He and heavier elements, at least at a given ener\-gy/nu\-cle\-on (the
more relevant parameter for acceleration, Paper II).  However, as we
show in Paper II, a shock can simultaneously accelerate He less
efficiently and heavy volatile elements more efficiently than H, if the
shock has a fairly low Mach number (e.g., $\sim 10$ or less) and all
elements have the same temperature in the unshocked medium.
\parano -- By contrast, there is only a very weak mass dependence of
the refractory element overabundances, or none at all.
It is well known, for instance, that the GCRS Fe/Mg ratio is close to
solar, enhanced by 20\% at the most.
The current ultra-heavy  abundance estimates relative to Fe suggest modest
excesses of most elements with $Z>40$, but these analysis need
confirmation ({\S}~2.1 and 4).
\parano -- Regarding the two intermediate classes of volatility, they
show intermediate overabundances and fit beautifully into the picture.
In particular, the low Na,P/Cu,Ga ratios in the ``semi-volatile" group,
the low S/Zn,Se ratio and the high ``Pb" abundance (if confirmed) among
the ``volatile" one (Fig.~5), are now readily interpreted in terms of
a mass effect.
\parano -- With the current errors, it is, of course, not possible to
know whether the ``volatile" elements behave significantly differently
from the ``highly volatile" ones, or not.
}   
\vskip 0.3truecm

Of course, the mass number $A$ is not a physical parameter capable of
governing by itself the acceleration efficiency for the various
elements.
The observed rough mass dependence of GCRS overabundances of the more
volatile elements most likely just reflects an actual correlation with
$A/Q$, \iec a rigidity dependence of the acceleration efficiency
(Paper~II).
In any ionization situation, indeed, $A/Q$ is, by and large, a
monotonically increasing function of $A$ (with local variations of this
generally monotonic increase related to the electronic shell
structure).
Clearly, the appropriate abscissa scale in Fig.~6 would have
been $A/Q$, rather than $A$.
But plotting an $A/Q$ scale would have required the knowledge of the
ionization states for all elements in the source gas, which would go
far beyond the scope of the present work
and will be investigated in a forthcoming paper.
We have, however, stressed this point in Fig.~6 by denoting the
abscissa scale ``$A \sim(A/Q)^{\alpha}$~", leaving $\alpha$
unspecified.

Qualitatively, we can say that the accelerated gas cannot be a purely
collisionally ionized gas around $\sim$~10$^{4}$~K, since in such a
gas Ne and He, for example, would be entirely neutral, hence not
accelerated.
It could be hot $\sim$~10$^{6}$~K gas, in which grains have been
somehow preserved, in which case the charge states $Q$ are a rather
smooth function of $A$,
and we get, very roughly, $A/Q \approx A^{0.4}$ (Arnaud \&
Rothenflug 1985).
It might also be $\sim$~10$^{4}$~K gas photoionized by stellar UV
radiation, in which case most elements will have charges of $Q$~=~+1
or +2,
so that $A/Q$~= (0.5 to 1) $\times A$;
the pure mass scale on Fig.~6 would then be relevant as an $A/Q$
scale, to within a left ward shift by a factor of 2 for the points
representing the elements with $Q$~=~+2, \iec those with a low second
ionization potential.
The somewhat low N/\net ratio (two elements with neighboring
masses) could be understood in this context, if N was predominantly in
the N$^{+2}$ state, and \net in the \neto$^{+1}$ state.
By contrast, whatever the charge state of He, \Heone or \Hetwoo, the
high H/He ratio cannot be understood in these terms, but it {\it can}
be understood as a direct effect of shock acceleration (Paper II).

Note that $A/Q$-dependent abundance enhancements similar to those
observed among the GCR volatiles are reported to exist in several
heliospheric accelerated particle populations:

\vskip 0.3truecm

\parano (i) Cummings \& Stone (1996) claim that the ``ano\-malous
cosmic rays'', accelerated by the solar wind termination shock in the
outer heliosphere, show a $A/Q$ enhancement, and attribute this to the
same effect of shock smoothing discussed in Paper II for GCR's.

\parano (ii) Smooth $A/Q$-dependent enhancements are clearly found in {\it
gradual\/} SEP events accelerated by coronal mass ejection associated
shocks in the corona and interplanetary medium (Mogro-Campero \&
Simpson 1972; Meyer 1985a; Breneman \& Stone 1985; Stone 1989; Garrard
\& Stone 1993; Reames 1995).  The $A/Q$ enhancements are found
superimposed upon the FIP-bias of the coronal and solar wind
composition relative to photosphere (Meyer 1993, 1996); this FIP-bias
actually accounts for part of the bulk heavy element enhancements
relative to photosphere noted early on (Mogro-Campero \& Simpson
1972).
Here, however, heavier, higher-$A/Q$ elements, while most frequently
enhanced, are also sometimes depleted relative to lighter ones.  The
depletions of heavier elements are mostly observed when the spacecraft
is poorly connected to the flare site, while the enhancements are
generally found in well connected events (Cane, Reames \& von
Rosenvinge 1991).
So, shock smoothing, which can produce only heavier element
enhancements (Paper II), is certainly not always the dominant factor
shaping the accelerated solar particle composition.  For instance,
acceleration by weaker, less smoothed shocks, in a parallel or a
perpendicular geometry, wave generation and saturation, variations in
shock geometry, particle trapping and escape, and contributions of
stochastic and resonant wave acceleration may all play important
roles.

\parano (iii) Note that the heavy element enhancements found in {\it
impulsive\/}, \heeeo-rich SEP events have quite different, specific
characteristics, and are to be explained in terms of resonant wave
acceleration (e.g.\ Reames, Meyer \& von Rosenvinge 1994; Steinacker
et al.\ 1997 and ref.\ therein).

\parano (iv) Shock acceleration at the earth bow shock has been
studied in detail with {\it in situ} spacecraft measurements and
comprehensive modeling (e.g., Jones \& Ellison 1991).  Clear evidence
for non-linear effects from efficient shock acceleration has been
reported, including the $A/Q$ enhancement of diffuse heavy ions
accelerated at the quasi-parallel portion of the shock (e.g., Ellison,
M\"obius, \& Paschmann 1990).  While the observed enhancements are
modest compared to those seen in GCR's, they are fully consistent with
non-linear shock acceleration theory and have not been successfully
explained by any alternative model.


\vskip12pt
\vbox {
\cl {6. SUMMARY AND CONCLUSIONS}
\vskip8pt

We have shown that the GCR source (GCRS) abundances of {\it all\/}
elements are best described (Fig.~6) in terms of
(i) a general enhancement of the refractory elements relative to the
volatile ones, and
(ii) among the volatile elements, an enhancement of the heavier
elements relative to the lighter ones;
this general trend accounts, in particular, for the well known low
abundances of H, He, and N in the GCRS;
besides C and O, for which a specific contribution is expected from
the Wolf-Rayet He-burning material component responsible for the \nett
excess,
only H, slightly enhanced relative to He at a given energy/nucleon,
does not entirely fit into this pattern;
this mass dependence most likely reflects a mass-to-charge ($A/Q$)
dependence of the acceleration efficiency.
Among the refractory elements, there is {\it no\/} such enhancement of
heavier species, or only a much weaker one.
}  

This conclusion is based on a detailed analysis of the GCRS
composition, in terms of both FIP and volatility.
In particular, the GCRS \ Na/Mg, Ge/Fe, Pb/Pt, and P/S ratios between
elements of comparable FIP and mass, but widely different volatilities,
are very difficult to interpret in terms of a FIP fractionation.
Specifically, a combination of a FIP and of a mass (or $A/Q$)
fractionation could not account for them (see further discussion in
Paper II).

We regard the strong similarity between the volatility-biased GCRS
composition and the FIP-biased composition of the solar corona, wind
and energetic particles as coincidental.  Note that this similarity
is, indeed, not complete: crucial elements such as Na and P ({\S}~4)
do seem to behave differently in GCR sources and SEP's, where they
clearly follow the FIP pattern (Garrard \& Stone 1993; Reames 1995).
By contrast, the $A/Q$-dependent enhancements found among the
GCR volatiles and in various heliospheric accelerated particle
populations should, in several instances, have the same causes.

To confirm  or disprove these views, new determinations of the GCRS
abundance of all low-FIP volatile and high-FIP {\it moderately\/}
volatile elements (in the lower left and upper right parts of Fig.~2)
would be essential.  In addition to the key elements already studied,
\iec
$_{11}$Na, $_{15}$P, $_{16}$S, $_{29}$Cu, $_{30}$Zn, $_{31}$Ga,
$_{32}$Ge, $_{34}$Se, $_{82}$Pb,
the following elements, whose GCRS abundances may become accessible in
the future, can serve as clues:
$_{17}$Cl,
$_{19}$K,
$_{25}$Mn,
$_{33}$As,
$_{35}$Br,
$_{37}$Rb,
$_{47}$Ag,
$_{48}$Cd,
$_{49}$In,
$_{50}$Sn,
$_{51}$Sb,
$_{52}$Te,
$_{55}$Cs,
$_{79}$Au,
$_{81}$Tl,
and
$_{83}$Bi.

In a companion paper (Ellison, Drury, \& Meyer 1997; Paper II),
this GCRS composition is interpreted in terms
of an acceleration of interstellar and/or circumstellar gas and dust
material by SNR shock waves.
Such shock waves, smoothed by the feedback pressure of the very
accelerated particles, preferentially inject and accelerate
the higher rigidity
ions.  Among the ISM gas-phase volatile elements, they therefore
enhance the higher $A/Q$, hence in practice the more massive elements
(early suggestions of this effect were given by Eichler 1979;
Ellison 1981; Ellison, Jones, \& Eichler 1981).
The same shock waves treat  the dust grains as extremely high $A/Q$
``ions" and  accelerate them very efficiently to $\sim$~0.1~MeV/n energies
where friction and sputtering become important.
The sputtered ions form a population of $\sim$~0.1~MeV/n refractory
elements which can be further accelerated by the shock, and for which
the
{\it crucial\/} early acceleration phases have
taken place while the ion was a member of the entire grain, hence
independent of its own individual mass.
So, both the {\it presence\/} of a strong mass dependence of the
abundance enhancements among the volatile elements, and its {\it
absence\/} among the refractories may be understood consistently.
Contrary to the earlier models in terms of FIP, such a picture
accelerates the GCR ions in a single step in a single site.
It also accounts naturally for the presence of a \netto-\cto-\os
excess in GCR's, since the shocks associated with the most massive
SNae accelerate their own pre-SN \netto-\cto-\oso-rich Wolf-Rayet wind
material.

\vskip18pt \noindent {Acknowledgments:} \ D. Ellison and L. Drury wish
to acknowledge the hospitality of the Service d'Astro\-phy\-sique,
Centre d'Etudes de Saclay where much of this work was carried out.  L.
Drury's visit was supported by the Commission of the European
Communities under contract ERBCHRXCT940604, and D. Ellison was
supported, in part, by the NASA Space Physics Theory Program.  The
authors also wish to thank
Monique Arnaud, Jean Ballet, Bob Binns, Michel Cass\'e, Anne
Decourchelle, Mike Duvernois, Jean-Jacques Engelmann, Philippe
Ferrando, Martin Israel, Frank McDonald, Georges Meynet, Renaud
Papoular, Etienne Parizot, Vladimir Ptuskin, Don Reames, Charles
Ryter, Blair Savage, Eun-Suk Seo, Christopher Sharp, Aim\'e Soutoul,
and Bill Webber
for valuable discussions regarding various aspects of this paper.


\vskip12pt
\vbox {
\cl {APPENDIX: THE $^{22}$Ne - $^{12}$C - $^{16}$O - RICH COMPONENT.}
\vskip8pt

There exists one important exception to the absence of signature of
specific nucleosynthetic processes in the GCRS composition ({\S}~2.2).
A large GCRS excess of the isotope \nett is derived from
observations.
The source ratio \nerat $\simeq  0.335\pm 0.065$
(Lukasiak et al.\ 1994; Duvernois et al.\ 1996; Webber, Lukasiak \&
McDonald 1997)
is enhanced by a factor of $\sim 4.4\pm 0.9$ relative to the low solar
reference ratio \nerato\sol $\simeq 0.076\pm 0.005$, on which
planetary Ne-B, solar wind, and recent SEP and local ISM values
derived from ``anomalous cosmic-ray" data now converge
(Anders \& Grevesse 1989; Selesnick et al.\ 1993; Mewaldt, Leske \&
Cummings 1996; and ref.\ therein).
}  

This large \nett excess, associated with the high GCRS C/O ratio
(1.7~$\times$ solar), seems a clear signature of the presence of a
He-burning material component in GCR's (Meyer 1981a, 1985b).
The absence of observed signatures of other, unrelated, specific
nucleosynthesis processes in the GCRS composition (\S 2.2), suggests an
origin of the material in Wolf-Rayet (WR) star wind material in which
pure He-burning material is being expelled into space during the WC
and WO stages, without other large anomalies
(Cass\'e \& Paul 1982; Maeder 1983; Prantzos, Arnould, \& Arcoragi
1987; Maeder \& Meynet 1993).

In massive stars such as WR's, the CNO cycle first turns the
entire initial CNO into $^{14}$N.
At the onset of He-burning this entire $^{14}$N in the He-burning
layer is briefly turned into $^{18}$O, which itself gets rapidly
turned into \netto.
The latter remains stable through most of the He-burning phase.
Meanwhile, the $^{4}$He gets progressively turned into $^{12}$C, and
later into $^{16}$O by addition of another $^{4}$He.
The $^{25,26}$Mg isotopes essentially start being formed only after
the He-burning phase, largely through \nett destruction.
In the case of WR stars, huge winds peel off the star to the point
where, first their N-rich H-burning zone (WN phase), then their C-, and
later O-rich He-burning core material is being blown off into space
(WC and WO stages)
(Prantzos et al.\ 1986; Prantzos, Arnould \& Arcoragi 1987; Maeder
1992; Schaller et al.\ 1992; Maeder \& Meynet 1994; Meynet et al.\
1994).

The time-averaged  \nett and $^{12}$C enhancements relative to solar
in the He-burning material can be roughly estimated from the {\it
initial\/} stellar abundances (Meyer 1981a, 1985b).
We now update the earlier estimates of these enhancements, based on
new GCRS and reference solar \nerat ratios ($\sim 1/2$
previous SEP ratios which were often used in earlier work).
As reference solar abundances, we adopt Grevesse et al.\ (1996)'s
elemental abundances and the above \nerat ratio, yielding H\sol
$\simeq 2.75\x{6}$, He\sol $\simeq  0.27\x{6}$, C\sol $\simeq  980$,
CNO\sol $\simeq 3280$, and
\netto\sol $\simeq  23.4$, on a  scale where Si\sol $\equiv 100$.
With these values, the conversion of all the initial CNO into $^{14}$N
and then into \nett yields a \nett enhancement by a factor of
CNO\solo/\netto\solo \ $\simeq  3280/23.4 \simeq 140$ in the He-burning layer.
In the same He-burning layer, the maximum possible $^{12}$C
enhancement is reached if all the initial H and He are converted into
$^{12}$C by the $3\alpha$ process.  This yields a maximal possible
$^{12}$C enhancement factor of (H\solo/4+He\solo)/3C\sol $\simeq
(2.75/4+0.27)\x{6}/(3\cdot980) \simeq 325$.  This value is, however,
not reached, because $^{12}$C starts being turned into $^{16}$O in the
later stages of the He-burning phase.
In addition, the $^{12}$C enhancement builds up progressively, so that
the averaged enhancement during the He-burning phase should be about
half the maximal value reached.  We will therefore adopt an averaged
$^{12}$C enhancement in the He-burning layer, of $\sim 35 \pm 10$\%
of the above maximal possible value, \iec of $\sim 115 \pm 33$.

To obtain the GCRS \nerat enhancement factor of $4.4\pm0.9$, the
He-burning component with \nerat enhancement of 140 must be highly
diluted in a component with solar \nerato, by a factor of
$140/(4.4\pm0.9-1) \simeq  44 \pm 12$.
With the same dilution factor, the predicted $^{12}$C excess in GCR
sources is $1+(115 \pm 33)/(44 \pm 12) \simeq 4.0 \pm 1.6$.
As a result, the carbon abundance in the main GCRS component, {\it
not\/} affected by WR He-burning nucleosynthesis, is $\sim 1/(4.0 \pm
1.6)$, i.e.\ $\sim 18$ to 42\% of the total GCRS carbon abundance.
This rough estimate has been plotted on Figs.~1, 5, and 6.
It is based on  WR nucleosynthesis properties only.
If, in addition, a fraction of the WC wind carbon gets preferentially
accelerated because it has condensed in solid carbons in the C-rich WC
star wind material ({\S}~3), the fraction of the observed GCRS carbon
originating in the main, {\it non\/}-WR, GCR component may be even
smaller.

Which accompanying composition anomalies should one expect?
At the onset of He-burning, $^{14}$N is first turned into $^{18}$O.
The smaller the stellar mass, the longer this $^{18}$O survives,
before being converted into \netto.  We might therefore find an
associated $^{18}$O excess,
{\it if\/} some stars are peeled off down to the He-burning core
before $^{18}$O has been destroyed.
%
Further, if WC type WR star wind material contributes to GCR's,
we expect similar contributions from the wind material of the
preceding and subsequent N-rich WN and O-rich WO phases.
The expected associated N and O excesses in GCR's, however, cannot be
evaluated with any certainty.
The WR star wind yields of N and O, relative to those of \nett and C,
indeed, depend critically upon the still unsettled time profile of the
mass loss rate as the star evolves, and upon the degree of mixing
between stellar layers (Maeder 1992; Maeder \& Meynet 1994; Meynet
1996).
In addition, the efficiency of wind material acceleration into GCR's
may also be a function of stellar mass.
In the WN phase, CNO-cycle H-burning material is being expelled in
the wind.  In this material, the N enhancement is equal to
CNO\solo/N\sol $\simeq  3280/257 \simeq  12.8$ only.
The actual expected N excess depends upon the relative strengths of
the winds in the WN and WC phases (\egc Maeder 1992, Maeder \& Meynet
1994).
If the WN and WC wind contributions are very roughly comparable, as
could be the case, this N-rich component, also diluted by a factor on
the order of $\sim 44$, yields a GCRS N excess by a factor of
$\sim 1+12.8/44 \simeq  1.29$ only.
In the WO phase, O-rich gas, due to the $^{12}$C $+\ \alpha\rightarrow
\ ^{16}$O reaction in the later stages of He-burning, is being
expelled in the Wolf-Rayet wind.
So, we  also expect an $^{16}$O excess.
For each stellar mass, the $^{16}$O yield in the wind depends
critically upon the degree of nuclear evolution of the He core at the
time its gets significantly peeled off by the winds.
Finally, the conversion of the \nett into $^{25,26}$Mg takes place
only very late in the He-burning phase, and can probably be neglected
in the WR wind material.
\footnote{$^1$}
{In a more precise, but more model-dependent, approach, we can
consider the calculated total yields of $^{12}$C, $^{14}$N, $^{16}$O,
and \nett from WR star winds, by integrating the wind contribution of
all WR stars over the stellar initial mass function (IMF).
Normalized to the above \nett excess of 140, Maeder (1992)'s study
leads to excess factors of 92, 16, and 8 for $^{12}$C, $^{14}$N, and
$^{16}$O, respectively.
A similar calculation by Meynet (1996),
assuming a twice higher mass loss rate during the MS and WNL phases
(Maeder \& Meynet 1994; Meynet \etal 1994, 1996), but still with
standard mixing, yields very similar figures for $^{12}$C, $^{16}$O,
and \netto, and a higher $^{14}$N yield.
These calculations confirm that $^{12}$C and \nett are enhanced by
comparable amounts.  The N yield is very model dependent.
The $^{16}$O excess of only $\sim $ 9\% of that of $^{12}$C is
uncomfortably low.  It does not allow a large fraction of the GCRS
oxygen to originate in WR star nucleosynthesis, which may preclude the
interpretation of the high GCRS O/\net ratio (Fig.~6) in terms of WR
star nucleosynthesis.  But different assumptions regarding the mass
loss rate and/or mixing are still to be explored.
}  

The GCRS isotope ratios derived from observations are consistent
with the above features: recent analysis suggest a low
$^{13}$C/$^{12}$C ratio (i.e.\ a $^{12}$C excess) and possibly a high
$^{18}$O/$^{16}$O ratio (yet to be confirmed), and indicate the
absence of a significant $^{25,26}$Mg/$^{24}$Mg enhancement
(Ferrando 1994; Lukasiak et al.\ 1994; Duvernois et al.\ 1996; Webber,
Lukasiak, \& McDonald 1997).

\vfill\eject


%
\def\aa#1#2#3{ 19#1, {\it A\&A,} {\bf #2}, #3.}
\def\aasup#1#2#3{ 19#1, {\it A\&AS,} {\bf #2}, #3.}

\def\annrev#1#2#3{ 19#1, {\it Ann.\ Rev.\ Astr.\ Ap.,} {\bf #2}, #3.}
\def\apj#1#2#3{ 19#1, {\it ApJ,} {\bf #2}, #3.}
\def\apjlet#1#2#3{ 19#1, {\it ApJ,} {\bf  #2}, #3.}

\def\apjs#1#2#3{ 19#1, {\it ApJS,} {\bf #2}, #3.}

\def\asr#1#2#3{ 19#1, {\it Adv.\ Space Res.,} {\bf #2}, #3.}
\def\araa#1#2#3{ 19#1, {\it Ann.\ Rev.\ Astr.\ Ap.,} {\bf #2},
   #3.}
\def\ass#1#2#3{ 19#1, {\it Astr.\ Sp.\ Sci.,} {\bf #2}, #3.}

\def\jpG#1#2#3{ 19#1, {\it J.\ Phys.\ G: Nucl.\ Part.\ Phys., }
     {\bf #2}, #3.}
\def\mnras#1#2#3{ 19#1, {\it MNRAS,} {\bf #2}, #3.}

\def\nucphysB#1#2#3{ 19#1, {\it Nucl.\ Phys.\ B (Proc.\ Suppl.),}
    {\bf #2}, #3.}

\def\ssr#1#2#3{ 19#1, {\it Space Sci.\ Rev.,} {\bf #2}, #3.}
\def\science#1#2#3{ 19#1, {\it Science,} {\bf #2}, #3.}

\def\RoySoc#1#2#3{ 19#1, {\it Phil.\ Trans.\ R.\ Soc.\ Lond.,} {\bf #2}, #3.}
\def\GCA#1#2#3{ 19#1, {\it Geochim.\ Cosmochim.\ Acta,} {\bf #2}, #3.}
\def\ps#1#2#3{ 19#1, {\it Physica Scripta,} {\bf #2}, #3.}
\def\icrcmunich#1#2{ 1975, in {\it Proc.\ 14th Int.\ Cosmic-Ray Conf.\
(Munich)}, {\bf #1}, #2.}

\def\icrckyoto#1#2{ 1979, in {\it Proc.\ 16th Int.\ Cosmic-Ray Conf.\
(Kyoto)}, {\bf #1}, #2.}
\def\icrcparis#1#2{ 1981, in {\it Proc.\ 17th Int.\ Cosmic-Ray Conf.\
(Paris)}, {\bf #1}, #2.}
\def\icrcbang#1#2{ 1983, in {\it Proc.\ 18th Int.\ Cosmic-Ray Conf.\
(Bangalore)}, {\bf #1}, #2.}

\def\icrcmoscow#1#2{ 1987, in {\it Proc.\ 20th Int.\ Cosmic-Ray Conf.\
(Moscow)}, {\bf #1}, #2.}
\def\icrcadel#1#2{ 1990, in {\it Proc.\ 21st Int.\ Cosmic-Ray Conf.\
(Adelaide)}, {\bf #1}, #2.}
\def\icrcdub#1#2{ 1991, in {\it Proc.\ 22nd Int.\ Cosmic-Ray Conf.\
(Dublin)}, {\bf #1}, #2.}
\def\icrccalgary#1#2{ 1993, in {\it Proc.\ 23rd Int.\ Cosmic-Ray Conf.\
(Calgary)}, {\bf #1}, #2.}
\def\icrcrome#1#2{ 1995, in {\it Proc.\ 24th Int.\ Cosmic-Ray Conf.\
(Rome)}, {\bf #1}, #2.}

{\parskip=0pt


\vskip12pt
\cl {REFERENCES}
\vskip8pt

\reff Anders, E., \& Grevesse, N., \GCA{89}{53}{197}

\reff Anders, E., \RoySoc{77}{A~285}{23}

\reff Arnaud, M., \& Cass\'e, M., \aa{85}{144}{64}

\reff Arnaud, M., \& Rothenflug, R., \aasup{85}{60}{425}

\reff Arnett, D., \araa{95}{33}{115}

\reff Arnould, M., \asr{84}{4}{(2-3), 45}

\reff Bibring, J.P., \& Cesarsky, C.J., \icrcparis{2}{289}

\reff Binns, W.R., \ 1995a, {\it Adv.\ Space Res.,} {\bf 15}, (6), 29.

\reff Binns, W.R., \ 1995b, in {\it Proc.\ 24th Int.\ Cosmic-Ray
Conf.\ (Rome)}, {\bf 3}, 384.

\reff Binns, W.R., Fickle, R.K., Garrard, T.L., Israel, M.H.,
Klarmann, J., Stone, E.C., \& Waddington, C.J.,
\apjlet{81}{247}{L118}                            

\reff Binns, W.R., Garrard, T.L., Gibner, P.S., Israel, M.H.,
Kertzman, M.P., Klarmann, J., Newport, B.J., Stone, E.C., \&
Waddington, C.J., \apj{89}{346}{997}              


\reff Breneman, H.H., \& Stone, E.C., \apjlet{85}{299}{L57}

\reff Byrnak, B., Lund, N., Rasmussen, I.L., Rotenberg, M., Engelmann,
J., Goret, P., \& Juliusson, E., \icrcbang{2}{29} 

\reff Cane, H.V., Reames, D.V., \& von Rosenvinge, T.T.,
\apj{91}{373}{675}

\reff Cardelli, J.A., \science{94}{265}{209}


\reff Cass\'e, M., \& Goret, P., \apj{78}{221}{703}

\reff Cass\'e, M., Goret, P., \& Cesarsky, C.J., \icrcmunich{2}{646}

\reff Cass\'e, M., \& Paul, J.A., \apj{82}{258}{860}

\reff Cesarsky, C.J., \& Bibring, J.P., \ 1981, in: {\it Origin of
Cosmic Rays\/}, IAU Symp.\ No.\ 94, G.\ Setti, G.\ Spada, \& A.W.\
Wolfendale eds., (Kluwer), p.\ 361.


\reff Clinton, R.R., \& Waddington, C.J., \apj{93}{403}{644}

\reff Connell, J.J., \& Simpson, J.A., \apjlet{97}{475}{L61}

\reff Cummings, A.C., \& Stone, E.C., \ssr{96}{78}{117}

\reff Draine, B.T., \ 1990, in: {\it The Evolution of the
Interstellar Medium\/}, L.\ Blitz ed., ASP Conf.\ Series Vol.\ 12
(Astr.\ Soc.\ of the Pacific), p.\ 193.

\reff Domingo, C., Font, J., Baixeras, C., \& Fern\'andez, F.,
\icrcrome{2}{572}

\reff Drake, J.J., Laming, J.M., \& Widing, K.G., \ 1997, {\it ApJ},
in press (March 20, 1997).

\reff Drury, L.O'C., \& Keane, A.J., \nucphysB{95}{39A}{165}

\reff DuVernois, M.A., Garcia-Munoz, M., Pyle, K.R., Simpson, J.A.,
\& Thayer, M.R., \apj{96}{466}{457}      

\reff DuVernois, M.A., Simpson, J.A., \& Thayer, M.R.,
\aa{96}{316}{555}                        
%

\reff DuVernois, M.A., \& Thayer, M.R., \apj{96}{465}{982}


\reff Dwek, E., 1979, Orange Aid Preprint OAP-570, Caltech.

\reff Dwek, E., 1997, submitted to {\it ApJ}.

\reff Eichler, D., \apj{79}{232}{106}

\reff Ellison, D.C., 1981, Ph.D. Thesis, The Catholic University  of
America.

\reff Ellison, D.C., Drury, L.O'C., and Meyer, J.P., 1997,
{\it ApJ\/}, in press [Paper II].

\reff Ellison, D.C., Jones, F.C., and Eichler, D., 1981, {\it
J.\ Geophys.\/}, {\bf 50}, 110.

\reff Ellison, D.C., M\"obius, E., \& Paschmann, G.,
\apj{90}{352}{376}

\reff Engelmann, J.J., Ferrando, P., Soutoul, A., Goret, P.,
Juliusson, E., Koch-Miramond, L., Lund, N., Masse, P., Peters, B.,
Petrou, N., \& Rasmussen, I.L., \aa{90}{233}{96}

\reff Engelmann, J.J., Goret, P., Juliusson, E., Koch-Miramond, L.,
Lund, N., Masse, P., Rasmussen, I.L., \& Soutoul, A. \aa{85}{148}{12}

\reff Epstein, R.I., \mnras{80}{193}{723}

\reff Feldman, U., \ps{92}{46}{202}

\reff Ferrando, P., \jpG{93}{19}{S53}

\reff Ferrando, P., \ 1994, in {\it Proc.\ 23rd Int.\ Cosmic-Ray
Conf.\ (Calgary), Invited, Rapporteur and Highlight Papers}, D.A.\
Leahy, R.B.\ Hicks, \& D.\ Venkatesan eds., (World Scientific), p.\
279.

\reff Forman, M.A., Ramaty, R., \& Zweibel, E.G., 1986, in {\it
Physics of the Sun,} Vol. II, D. Reidel Publ.\ Co., Dordrecht, Holland.

\reff Fowler, P.H., Walker, R.N.F., Masheder, M.R.W., Moses, R.T.,
Worley, A., \& Gay, A.M., \apj{87}{314}{739}

\reff Garrard, T.L., Israel, M.H., Klarmann, J., Stone, E.C.,
Waddington, C.J., \& Binns, W.R., \icrcadel{3}{61}

\reff Garrard, T.L., \& Stone, E.C., \icrccalgary{3}{384}

\reff Geiss, J., Gloeckler, G., \& von Steiger, R., \ssr{95}{72}{49}

\reff Grevesse, N., Noels, A., \& Sauval, A.J., \ 1996, in: {\it
Cosmic Abundances\/}, S.S.\ Holt \& G.\ Sonneborn eds., ASP Conf.\
Series Vol.\ 99 (Astr.\ Soc.\ of the Pacific), p.\ 117.

\reff H\'enoux, J.C., \asr{95}{15}{(7), 23}


\reff Israel, M.H., \ 1996, private communication.

\reff Jones, F.C., \&  Ellison, D.C., \ssr{91}{58}{259}

\reff Joseph, C.L., \apj{88}{335}{157}

\reff Jun, B.-I., \& Norman, M.L., \apj{96}{472}{245}

\reff Laming, J.M., Drake, J.J., \& Widing, K.G., \apj{96}{462}{948}

\reff Leske, R.A., \apj{93}{405}{567}

\reff Leske, R.A., \& Wiedenbeck, M.E., \icrccalgary{1}{571}

%

\reff Lukasiak, A., Ferrando, P., McDonald, F.B., \& Webber, W.R.,
\apj{94}{426}{366}   

\reff Lukasiak, A., McDonald, F.B., Webber, W.R., \& Ferrando, P.,
\ 1997, {\it Adv.\ Space Res.,} in press (Birmingham COSPAR Paper
E1.7-0004).

\reff Lund, N., \asr{84}{4}{(2-3), 5}    

\reff Maeder, A., \aa{83}{120}{130}                    

\reff Maeder, A., \aa{92}{264}{105}                    

\reff Maeder, A., \& Meynet, G., \aa{93}{278}{406}    

\reff Maeder, A., \& Meynet, G., \aa{94}{287}{803}    


\reff Mathis, J.S., \apj{96}{472}{643}

\reff Mewaldt, R.A., Leske, R.A., \& Cummings, J.R., \ 1996, in: {\it
Cosmic Abundances\/}, S.S.\ Holt \& G.\ Sonneborn eds., ASP Conf.\
Series Vol.\ 99 (Astr.\ Soc.\ of the Pacific), p.\ 381.

\reff Meyer, J.P., \ 1981a, in {\it Proc.\ 17th Int.\ Cosmic-Ray
Conf.\ (Paris)}, {\bf 2}, 265.                         

\reff Meyer, J.P., \ 1981b, in {\it Proc.\ 17th Int.\ Cosmic-Ray
Conf.\ (Paris)}, {\bf 2}, 281.                         

\reff Meyer, J.P., \apjs {85a}{57}{151}

\reff Meyer, J.P., \apjs {85b}{57}{173}

\reff Meyer, J.P., \ 1988, in: {\it Origin and Distribution of the
Elements\/}, G.J.\ Mathews ed., (World Scientific), p.\ 310.

\reff Meyer, J.P., \ 1993, in: {\it Origin and Evolution of the
Elements\/}, N.\ Prantzos, E.\ Vangioni-Flam \& M.\ Cass\'e eds.,
(Cambridge Univ.\ Press), p.\ 26; or {\it Adv.\ Space Res.,} {\bf
13}, (9), 977.

\reff Meyer, J.P., \ 1996, in: {\it The Sun and Beyond\/}, L.\
Celnikier \& J.\ Tr\^an Thanh V\^an eds., (Gif-sur-Yvette: Editions
Fronti\`eres), p.~27; or in: {\it Cosmic Abundances\/},
S.S.\ Holt \& G.\ Sonneborn eds., ASP Conf.\ Series Vol.\ 99 (Astr.\
Soc.\ of the Pacific), p.\ 127.

\reff Meyer, J.P., Cass\'e, M., \& Reeves, H., \icrckyoto{12}{108}

\reff Meynet, G., \ 1996, private communication.

\reff Meynet, G., Arnould, M., Prantzos, N., \& Paulus, G.,\ 1996,
{\it A\&A}, in press.

\reff Meynet, G., Maeder, A., Schaller, G., Schaerer, D., \&
Charbonnel, C., \aasup{94}{103}{97}

\reff Mogro-Campero, A., \& Simpson, J.A., \apjlet{72}{177}{L37}

\reff O'Sullivan, D., Thompson, A., Wentzel, K.P., \& Jansen, F.,
\asr{95}{15}{(6), 25}

\reff Papini, P., et al., \icrccalgary{1}{579}

\reff Prantzos, N., Arnould, M., \& Arcoragi, J.P.,
\apj{87}{315}{209}                    

\reff Prantzos, N., Cass\'e, M., \& Vangioni-Flam, E., \ 1993, in {\it
Origin and Evolution of the Elements\/}, N.\ Prantzos, E.\ Vangioni-Flam
\& M.\ Cass\'e eds., (Cambridge U.\ Press), p.~156.

\reff Prantzos, N., Doom, C., Arnould, M., \& De Loore, C.,
\apj{86}{304}{695}                    

\reff Ptuskin, V.S., \& Soutoul, A., \aa{90}{237}{445}

\reff Reames, D.V., \asr{95}{15}{(7), 41}

\reff Reames, D.V., Meyer, J.P., \& von Rosenvinge, T.T.,
\apjs{94}{90}{649}


\reff Reynolds, S.P., \& Ellison, D.C., \icrccalgary{2}{227}

\reff Sakurai, K., \asr{95}{15}{(6), 59}

\reff Savage, B.D., \& Sembach, K.R., \annrev{96}{34}{279}

\reff Schaller, G., Schaerer, D., Meynet, G., \& Maeder, A.,
\aasup{92}{96}{269}

\reff Selesnick, R.S., Cummings, A.C., Cummings, J.R., Leske, R.A.,
Mewaldt, R.A., Stone, E.C., \& von Rosen\-vinge, T.T.,
\apjlet{93}{418}{L45}                        

\reff Seo, E.S., 1997, \ private communication.

\reff Seo, E.S., \& Ptuskin, V.S., \apj{94}{431}{705}

\reff Seo, E.S., Ormes, J.F., Streitmatter., R.E., Stochaj, S.J., Jones, W.V.,
Stephens, S.A., \& Bowen, T., \apj{91}{378}{763}

\reff Seo, E.S., et al., \icrcrome{2}{648}

\reff Shibata, T., 1995, Rapporteur Talk, in {\it Proc.\ 24th Int.\
Cosmic-Ray Conf.\ (Rome)}, in press.

\reff Silberberg, R., \& Tsao, C.H., \apjlet{90}{352}{L49}

\reff Singh, K.P., White, N.E., \& Drake, S.A., \apj{96}{456}{766}

\reff Soutoul, A., Cesarsky, C.J., Ferrando, P., \& Webber, W.R.,
\icrcdub{2}{408}


\reff Steinacker, J., Meyer, J.P., Steinacker, A., \& Reames,
D.V., \apj{97}{476}{403}

\reff Stone, E.C., \ 1989, in {\it Cosmic Abundances of
Matter\/}, C.J.\ Waddington ed., AIPJConf.\ Proc.\ 183, (The Amer.\
Inst.\ of Physics), p.~72.

\reff Swordy, S., \ 1994, in {\it Proc.\ 23rd Int.\ Cosmic-Ray
Conf.\ (Calgary), Invited, Rapporteur and Highlight Papers}, D.A.\
Leahy, R.B.\ Hicks, \& D.\ Venkatesan eds., (World Scientific), p.\
243.

\reff Swordy, S., et al., \icrcrome{2}{652}

%
\reff Tarafdar, S.P., \& Apparao, K.M.V., \ass{81}{77}{521}

\reff Timmes, F.X., Woosley, S.E., \& Weaver, T.A., \apjs
{95}{98}{617}

\reff Tueller, J., Love, P.L., Israel, M.H., \& Klarman, J.,
\apj{79}{228}{582}

\reff Tylka, A.J., Adams, J.H., Beahm, L.P., Benning, L.S., Kleis, T.,
\& Witt, R.A., \icrcrome{2}{580}

\reff van der Hucht, K.A., \& Hidayat, B.\ (eds.), 1991, {\it
Wolf-Rayet Stars and Interrelations with Other Massive Stars in the
Galaxy\/}, IAU Symp.\ No.\ 143 (Kluwer, Dordrecht).

\reff van der Hucht, K.A., \& Williams, P.M.\ (eds.), 1995, {\it
Wolf-Rayet Stars: Binaries, Colliding Winds, Evolution\/}, IAU Symp.\
No.\ 163 (Kluwer, Dordrecht).

\reff Waddington, C.J., \apj{96}{470}{1218}

\reff Waddington, C.J., \ 1997, {\it Adv.\ Space Res.,} in press
(Birmingham COSPAR Paper E1.7-0012).


\reff Wasson, J.T., \ 1985, {\it Meteorites: Their Record of Early
Solar System History\/}, (New York: Freeman).

\reff Wasson, J.T., \& Kallemeyn, G.W., \RoySoc{88}{A~325}{535}

\reff Webber, W.R., Golden, R.L., \& Stephens, S.A.,
\icrcmoscow{1}{325}

\reff Webber, W.R., Lukasiak, A., \& McDonald, F.B., \ 1997, {\it
ApJ}, in press.

\reff Webber, W.R., \& McDonald, F.B., \apj {94}{435}{464}

\reff Woosley, S.E., \& Weaver, T.A., \apjs {95}{101}{181}

}  

\vfill\eject


\vskip12pt
\cl {FIGURE CAPTIONS:}
\vskip8pt

\vskip12pt
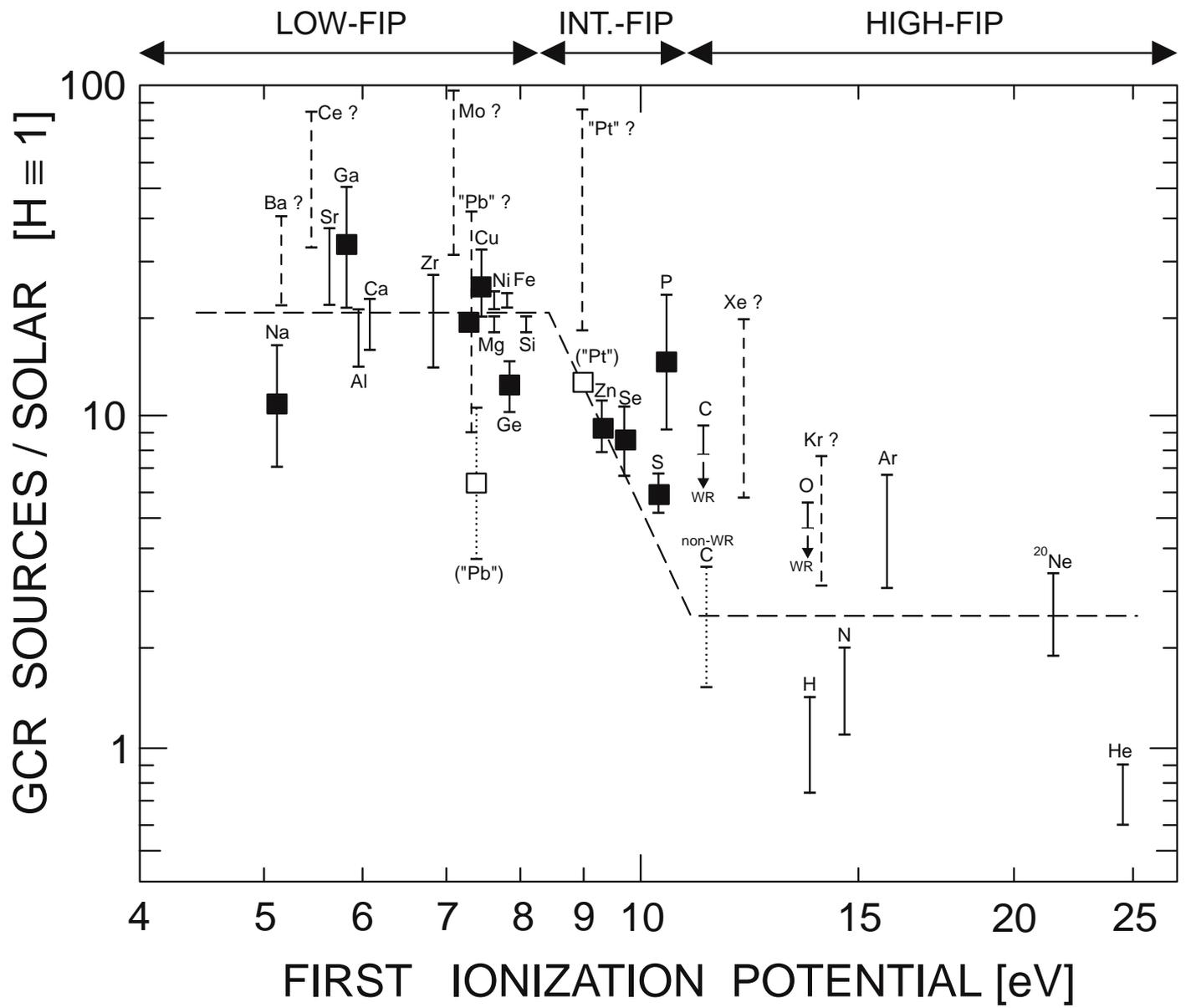
\noindent {\bf Fig.\ 1.} \
The standard GCRS to Solar abundance ratio versus FIP pattern in the
$\sim$~GeV range (ref.\ in {\S}~2.1).
The solar system reference abundances are from Grevesse et al.\
(1996), mainly from the meteoritic determinations, and are normalized
to H, at a given energy/nucleon.
Those elements which can serve as clues to distinguish between FIP and
condensation temperature \Tc as the parameter governing the GCRS
composition are emphasized by a solid square; their abundance
determination is discussed in more detail in {\S}~4.
The points for C and O are plotted as upper limits, in order to stress
that their total source abundance includes a specific \ct and \os
contribution associated with the \netto -rich component from
Wolf-Rayet stars; for \cto, we propose a tentative estimate of its
{\it non\/}-Wolf-Rayet source abundance (Appendix; we assumed
no preferential acceleration of C relative to \nett in the Wolf-Rayet
component).
For the same reason, the isotopic \net abundance has been plotted,
rather than that of the total elemental Ne.
We have marked by a dashed bar and a ``?" sign those ultra-heavy
elements whose source abundance {\it relative to Fe\/} is quite
uncertain, because they are very sensitive to the propagation model
and the spallation cross sections, so that a realistic uncertainty is
not easy to determine {\S}~2.1 and 4): all elements with $Z>40$, as
well as those whose estimated primary fraction is $<50$\% ($_{36}$Kr,
$_{42}$Mo, $_{54}$Xe; Binns 1995).
This is, in particular, the case for the crucial ``Pt"- and ``Pb"-group
elements with $Z$ = 74 -- 80 and 81 -- 83, whose source abundances
relative to Fe are highly uncertain.
But the source ``Pb"/``Pt" ratio is much better determined.
To visualize it, we have also plotted the ``Pb" point relative to the
``Pt" point {\it arbitrarily\/} placed where it would fit assuming the
standard FIP pattern (open squares).
Some elements with large error bars, which would unnecessarily confuse
the picture, have been omitted ($_{19}$K, $_{27}$Co, $_{50}$Sn,
$_{52}$Te).
In this figure, we define {\it ``low-FIP"\/}, {\it
``intermediate-FIP"\/}, and {\it ``high-FIP"\/} elements with FIP values
$<$ 8.5, 8.5 -- 11, and $>$ 11 eV, respectively.

\vskip12pt
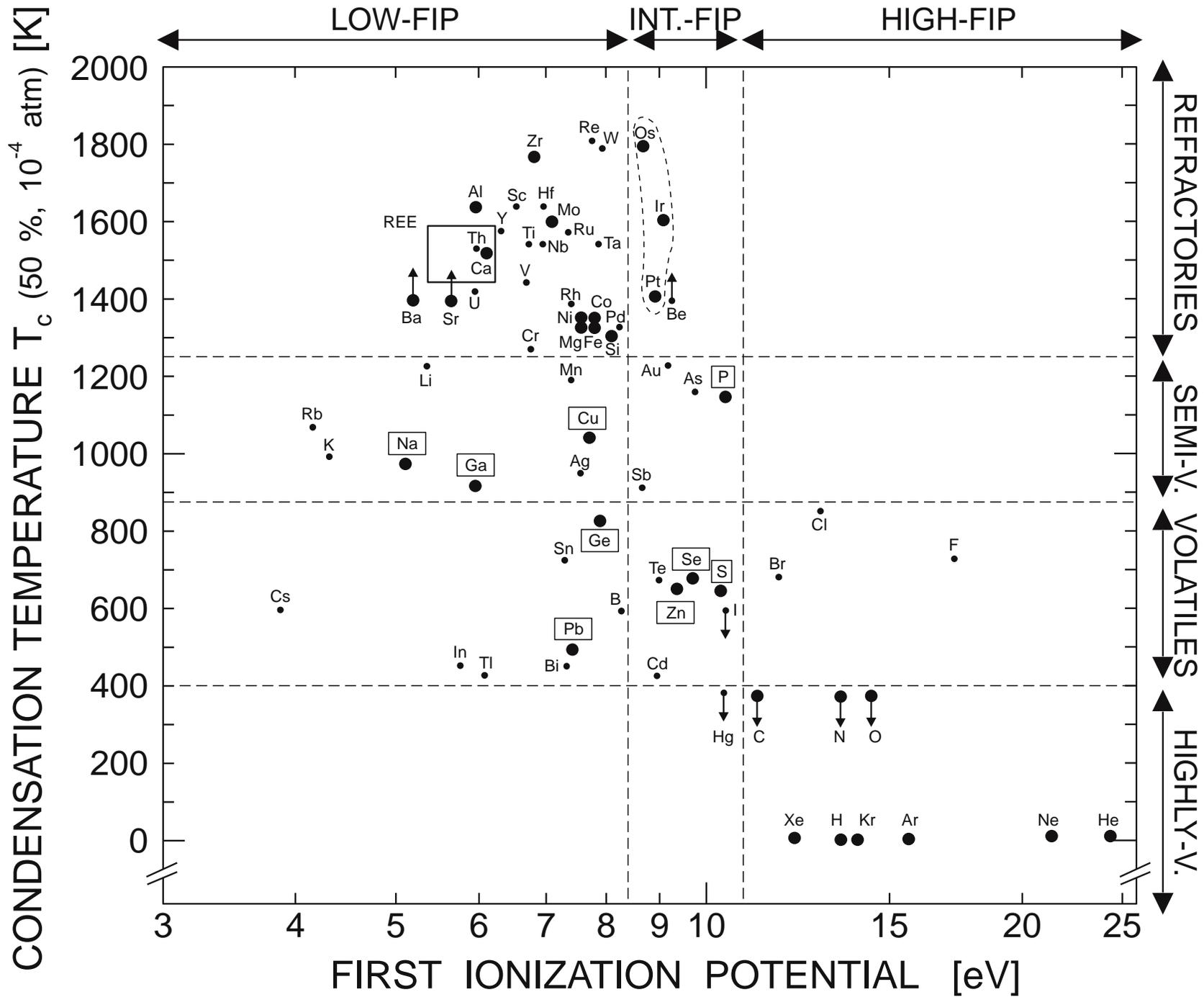
\noindent {\bf Fig.\ 2.} \
Cross plot of the condensation temperature \Tc of each chemical
element, versus the element's FIP.
Each $T_{c}$ value is the 50\% condensation temperature
of the dominant solid compound formed by the element,
for a solar initial gas composition at $10^{-4}$~atm.\
(from Wasson 1985, with additional information from Anders
1977 and Anders \& Grevesse 1989).
Along with the grouping of the elements into three groups according
to their FIP (Fig.~1), we define four classes of volatility:
{\it ``refractories"\/} with \Tc $>$ 1250~K,
{\it ``semi-volatiles"\/} with 1250~K $>$ \Tc $>$ 875~K,
{\it ``volatiles"\/} with 875~K $>$ \Tc $>$ 400~K,
and
{\it ``highly-volatiles"\/} with 400~K $>$ \Tc
(see also Figs.~3 and 4).
This figure shows the general anti-correlation between FIP and \Tco,
most lower-FIP elements being refractory, and higher-FIP ones
volatiles.
Those elements for which we currently have reasonably accurate
estimates their GCRS abundance are denoted by big, solid dots.
Among them, the elements lying outside the main FIP -- \Tc
correlation, which can serve as clues to distinguish between FIP and
\Tc as the parameter governing the GCRS composition, are framed (marked
by solid dots in Figs.~1, 3, and 5).
``REE" stands for ``rare earth elements".

\vskip12pt
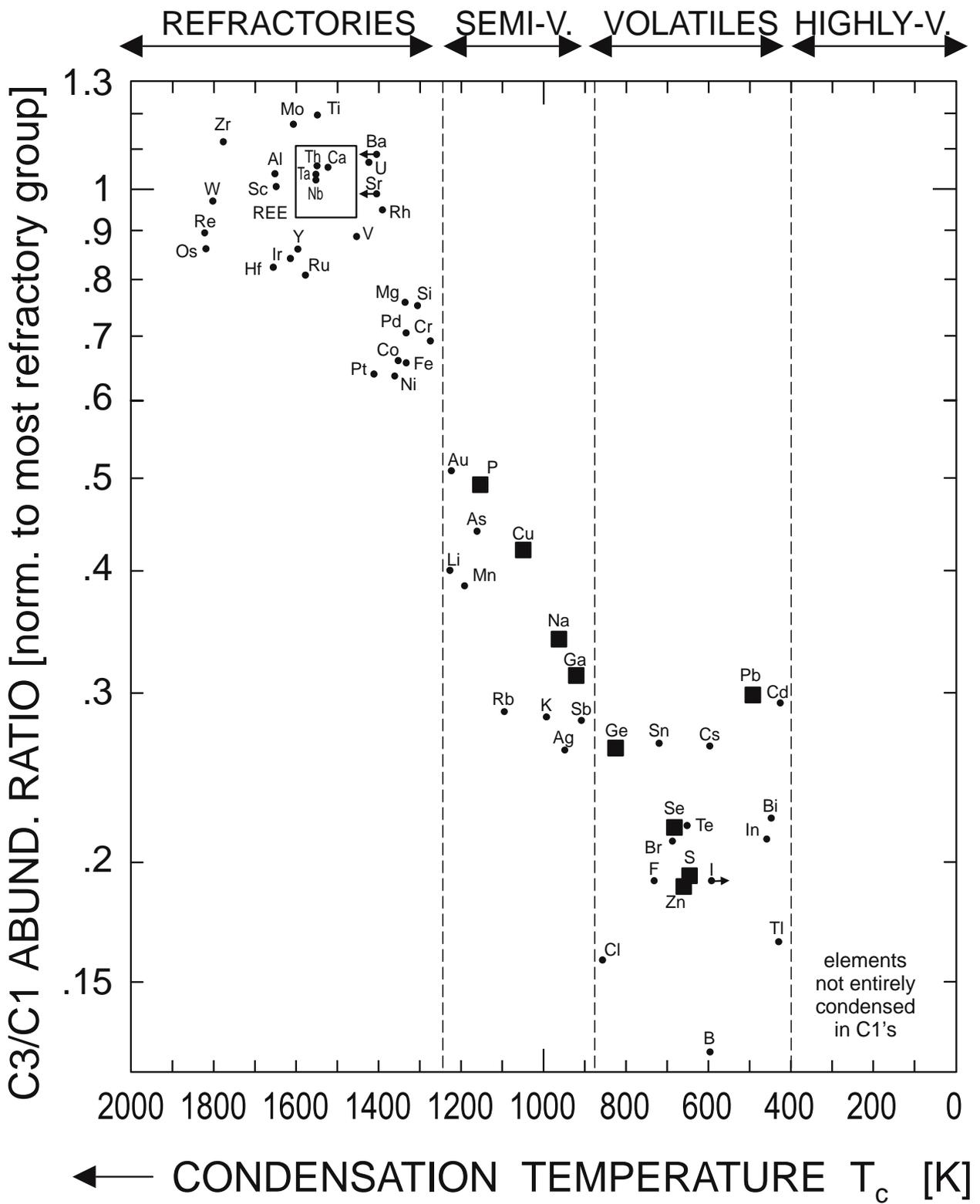
\noindent {\bf Fig.\ 3.} \
Depletion of the more volatile elements among the various types of
Carbonaceous Chondrites, illustrated by the Vigarano-type C3/C1
abundance ratio, versus condensation temperature \Tc (Wasson and
Kallemeyn 1988).
In C3's the more volatile elements are incompletely condensed, while
in C1's most elements are entirely condensed, with relative
abundances equal to those in the protosolar nebula.
All abundances are
normalized to the group of the most refractory elements.  REE stands
for ``rare earth elements".  The key elements for our analysis of GCR's
have been singled out by a solid square.  This figure shows that the
correlation between the C3/C1 abundance ratio and \Tc is quite good,
thus confirming the relevance of the parameter \Tco, at least in this
context.
It shows a few distinct groups of elements: (i)~very refractory,
lithophile elements with $T_{c} \gappeq 1400$~K and a group of
elements condensing as silicates or together with metallic Fe
(siderophiles) around $\sim$~1350~K, here together denoted {\it
``refractories"\/} with \Tc $>$ 1250~K;
(ii)~a group of {\it ``semi-volatile"\/} elements with 1250~K $>$ \Tc
$>$ 875~K, whose depletion in C3's varies rapidly with \Tco;
(iii)~a group of {\it ``volatile"\/} elements with 875~K $>$ \Tc $>$
400~K, with C3/C1 ratios $<$ 0.30;
and (iv)~the {\it ``highly-volatile"\/} elements, not significantly
condensed even in C1's, not plotted in the figure are H, C, N, O, and
the noble gases.

\vskip12pt
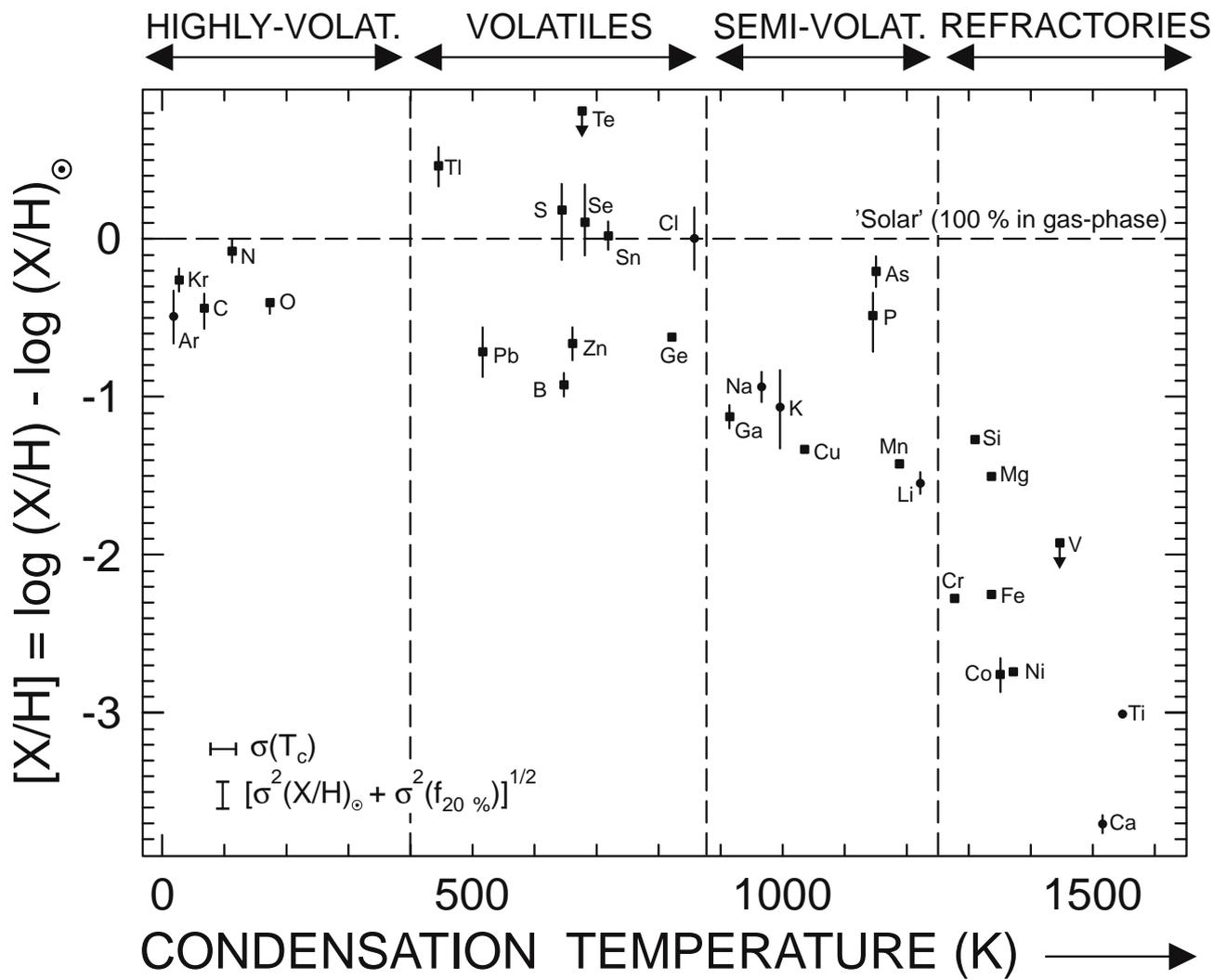
\noindent {\bf Fig.\ 4.} \
Elemental depletions relative to solar abundances in the ISM gas-phase
along the line of sight of $\zeta$~Oph (adapted from Savage and
Sembach 1996).  Our adopted four classes of volatility have been
singled out, as in Fig.~3.
The general trend is clearly a larger depletion of the more
refractory elements in the ISM gas-phase.
But the spread in the depletions between elements with similar values
of \Tc is much larger than in Fig.~3, presumably due to grain
destruction and reprocessing in the ISM. Note in particular the
apparent small depletion of comparatively refractory P (and As, which
has similar chemical properties).
The behavior of the more volatile elements suggests that some of the
spread may be observational, and that there exists some problems with
the solar normalization.

\vskip12pt
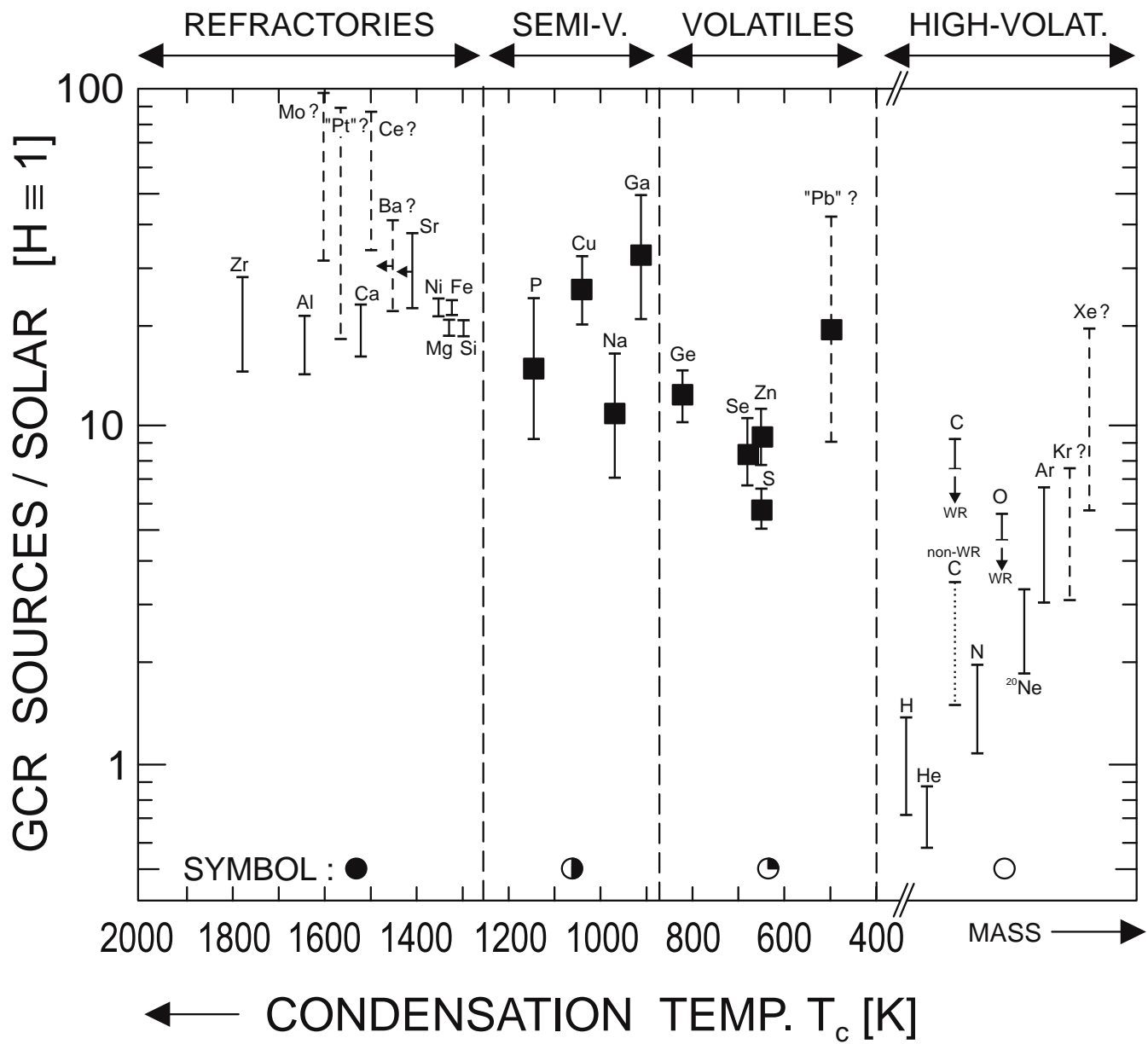
\noindent {\bf Fig.\ 5.} \
The same GCRS to Solar abundance ratios as in Fig.~1, this time
plotted versus condensation temperature \Tco.
See Fig.~1 caption; in particular, we have also singled out by a
solid square the clue elements for choosing between FIP and \Tc as the
relevant parameter controlling the GCRS composition; they all belong
to the two intermediate classes of volatility.
Clearly, the enhancements progressively decrease with decreasing \Tco,
throughout the four classes of volatilities.
For the ``highly-volatile" elements, \Tc doesn't make sense, and we
have plotted these elements simply in order of increasing mass;
%
except for H, their enhancements seem a monotonic function of the
mass
%
(for C, see Fig.~6 caption).
The large spread in the enhancements of the elements in the two
intermediate classes of volatility will also be interpreted as a mass
effect (Fig.~6).
The symbols used in  Fig.~6 to denote the elements in the
four classes of volatility are shown at the bottom of the figure.

\vskip12pt
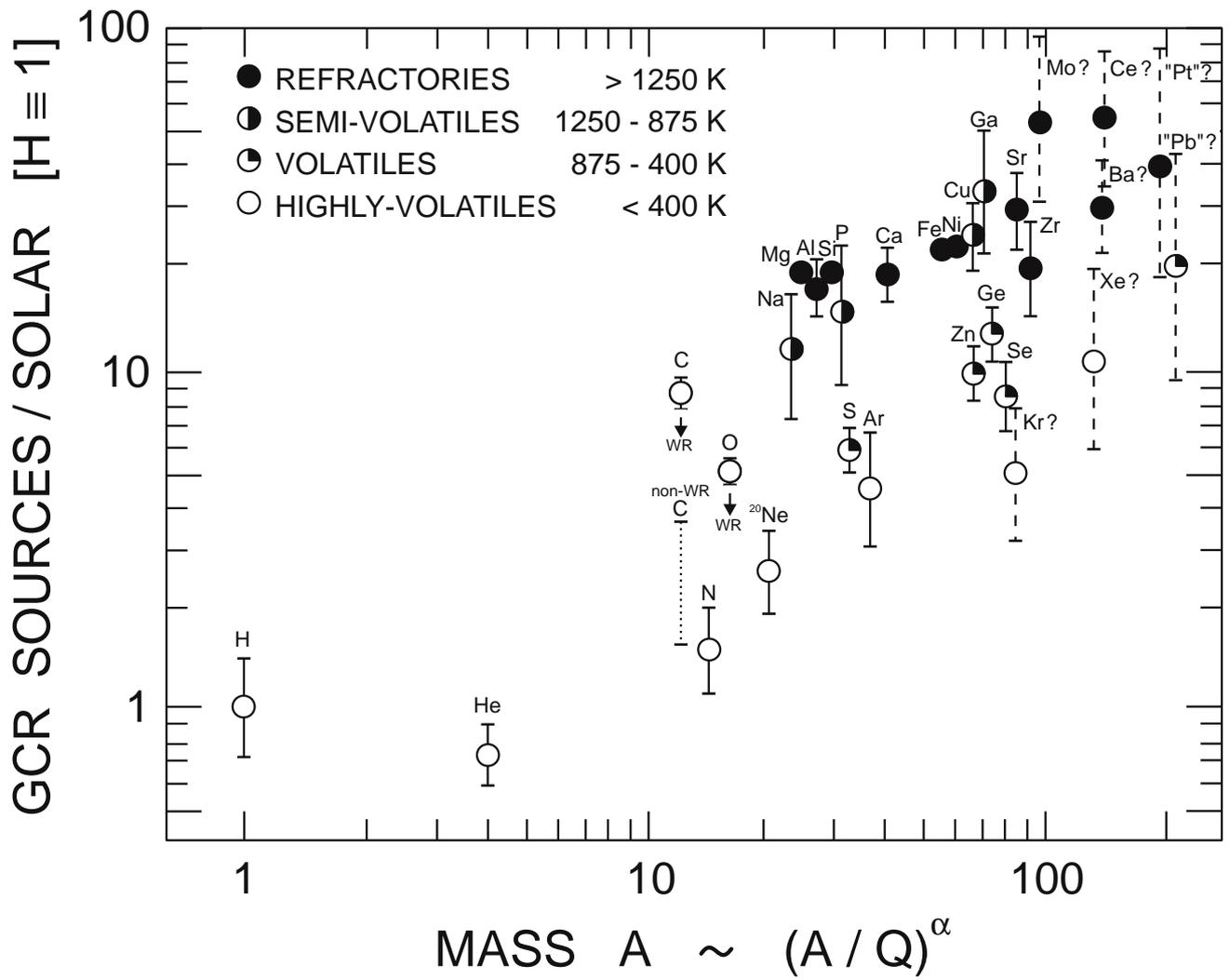
\noindent {\bf Fig.\ 6.} \
The same GCRS to Solar abundance ratios as in Figs.~1 and 5,
this time plotted versus element atomic mass number
$A$, for the elements in the
four classes of volatility.  See Fig.~1 caption.
Clearly, the more refractory elements are generally more enhanced than
the more volatile ones.
For the ``highly-volatile" elements, the GCRS/Solar enhancements seem
roughly $\propto A^{0.8\pm0.2}$, except for H, which poses a specific
problem;
(recall that our non-WR C estimate may still be an overestimate, if a
significant fraction of the C is locked in grains in the C-rich WC
wind material, and hence preferentially accelerated relative to \netto
; see {\S}{\S}~3, 5 and Appendix).
Physically, this apparent correlation of the enhancements with $A$
most likely reflects a correlation with $A/Q$ in the source gas, which
is a roughly monotonic function of $A$ in all practical ionization
situations.
For the ``refractory" elements, by contrast, there is only a very weak
increase of the enhancements with mass $A$, if any.
These contrasting behaviors will be interpreted in terms of the
volatile elements being accelerated as individual ions directly out of
the gas-phase, while the refractory elements are first accelerated as
constituents of entire grains.
The elements in the two intermediate classes of volatility show
intermediate behaviors.


\bye